\documentclass[print]{ati}	
\usepackage{graphicx,times}
\usepackage{lmodern}
\usepackage[sort&compress, numbers,super]{natbib} 
\bibpunct[,]{[}{]}{,}{n}{}{,}			
\usepackage{url}
\usepackage[colorlinks=true,breaklinks=true, linkcolor=red,urlcolor=magenta,citecolor=blue,anchorcolor=green]{hyperref}
\usepackage{caption}
\usepackage{rotating}
\usepackage{CJKutf8}	
\usepackage{fancyhdr}
\usepackage{subcaption}
\usepackage{booktabs}
\usepackage{amssymb}
\renewcommand{\citet}[1]{\citeauthor{#1}(\citeyear{#1})\cite{#1}}	

\immediate\write18{texcount -inc -sum ati-sample.tex > wordcount.txt}
   
\begin{document}

   \title{
   Initial Performance Tests of MASTA for Wide-Field Narrowband Imaging
}


   \author{Mingxuan Xu
      \inst{1,2}
      \ORCID{0009-0008-1918-9233}		
   \and Jiang-Tao Li\correspondingAuthor{}
     \inst{1}
     \ORCID{0000-0001-6239-3821}
   \and Li Ji
      \inst{1}
      \ORCID{0000-0001-7500-0660}
   \and Chen Zhang
      \inst{1}
   \and Yantian Zhu
      \inst{1,2}
   }
\correspondent{Jiang-Tao Li}	
\correspondentEmail{pandataotao@gmail.com}

\institute{Purple Mountain Observatory, Chinese Academy of Sciences, Nanjing 210008, China;
          \and
              School of Astronomy and Space Science, University of Science and Technology of China, Hefei 230026, China;
   }
   \date{Received:~August 2, 2026;   Accepted:~September 25, 2026;  Published Online:~ 2026; 
   \DOI{ati0000000} }			
   \citeinfo {Xu, M.-X. et al.}\volume{0}\issue{0} \pages{0--0}	
   \StartPage{0} 			
   \MonthIssue{October}		
   \copyrights {2026}     		
   \abstract{
Wide-field narrowband imaging provides an efficient approach for mapping low-surface-brightness line emission over large angular scales. We present initial performance tests of Multi-Application Survey Telescope Array (MASTA) for narrowband imaging. Using commissioning data obtained with the QHY461 and COSMOS-8K cameras, we characterize the basic reduction procedure, photometric calibration, image quality, and point-source detection limits. The measured surface-brightness and point-source detection limits become deeper with increasing exposure time, roughly following the expected background-limited scaling. Under the current observing and reduction conditions, the COSMOS-8K camera generally reaches deeper surface-brightness limits, particularly in the red filters, while QHY461 has smaller pixel size and provides better sampling of the telescope point spread function (PSF). As demonstrations, we perform continuum subtraction for the nearby galaxy M106 and the supernova remnant Cygnus Loop using adjacent narrowband filters and reveal the H$\alpha$ and [\ion{O}{III}] structures in M106 and filamentary structures in the Cygnus Loop. We further compare the MASTA surface brightness H$\alpha$ map of M106 with an independent SITELLE measurement. After accounting for [\ion{N}{II}] contamination and the wavelength-dependent response of the 6544\,\AA\ filter in the $f/1.7$ converging beam, the corrected median ratio is $0.973$. These tests indicate that MASTA is a promising facility for wide-field narrowband imaging. Further improvements in calibration frames, including flux calibration, field-dependent bandpass characterization, and spatially varying PSF modeling and matching will be required for quantitative studies of diffuse ionized gas.
  }

   \authorrunning{ASTRONOMICAL TECHNIQUES \& INSTRUMENTS }   
   \titlerunning{Xu M.-X. et al.: ~Prepare a LaTeX Manuscript for ATI }  
   \maketitle
   \setcounter{page}{\Page}	
%
%
\section{Introduction}
\label{sect:intro}

Narrowband imaging is an efficient observational technique for isolating emission-line radiation from the continuum background. It has been widely used to study star formation in galaxies~\citep{2012ARA&A..50..531K}, diffuse ionized gas in galactic disks and halos~\citep{2019ApJ...881...26V, 2023MNRAS.519.6098L}, supernova remnants~\citep{2012A&ARv..20...49V}, galactic outflows~\citep{2024ApJ...967...78L}, and filamentary ionized gas in galaxy groups and clusters~\citep{2008ApJ...687L..69K, 2008Natur.454..968F}. Compared with spectroscopy, narrowband imaging has lower spectral resolution but can provide much higher mapping efficiency over extended regions. It is therefore particularly useful for identifying low-surface-brightness line-emitting structures and for providing target selection and environmental context for follow-up spectroscopic observations.

Despite these advantages, wide-field narrowband imaging remains technically challenging. Most astronomical narrowband filters are interference filters, for which the effective central wavelength and bandpass depend on the angle of incidence of the incoming light. Rays entering the filter at larger incidence angles experience a shorter effective transmitted wavelength, producing a blueward shift of the passband; in a fast or non-telecentric optical system, the finite cone of rays can also broaden and distort the transmission curve~\citep{2000ASPC..195...34B, 2022SPIE12188E..21B}. As a result, the effective bandpass may vary across the field of view (FOV), limiting the usable field for very narrow filters and complicating accurate photometry of emission-line sources. This effect is especially important for large-field and fast optical systems, where the range of incidence angles is intrinsically large.

Another practical limitation is the correction of large-scale illumination patterns. Vignetting, scattered light, ghost reflections, and sky-background gradients can leave residual structures after standard flat-field calibration. These residuals are often small compared with the sky background but can be critical for studies of very diffuse and spatially extended emission. This is particularly relevant for observations of low-surface-brightness structures around nearby galaxies~\citep{2023MNRAS.519.6098L}, including extraplanar diffuse ionized gas, tidal features, and possible circumgalactic emission~\citep{2008ApJ...687L..69K, 2008Natur.454..968F, 2019ApJ...881...26V}. The difficulty of controlling large-scale systematics has motivated dedicated low-surface-brightness instruments and observing strategies, such as the Dragonfly Telephoto Array and its narrowband/spectral-line extensions~\citep{2014PASP..126...55A, 2020SPIE11445E..27L, 2022ApJ...927..136L}.

The choice of filter bandwidth is therefore a compromise among line contrast, throughput, field coverage, and redshift coverage. A narrower filter suppresses more continuum and sky background and can increase the contrast of a targeted emission line. However, it is also more sensitive to angle-dependent wavelength shifts, provides a smaller effective redshift interval, and may be less suitable for wide-field surveys or for sources with uncertain velocities. A somewhat broader narrowband filter can be preferable for survey applications, because it offers greater tolerance to field-dependent bandpass variations, peculiar velocities, and redshift uncertainties, while still providing substantially higher line contrast than broadband imaging.

Multi-Application Survey Telescope Array (MASTA) is a wide-field fast focal ratio optical telescope. Although it is originally designed mainly for broadband time-domain and survey observations, a dedicated set of narrowband filters has been developed to extend its capability to wide-field emission-line imaging. The fast focal ratio and large FOV make it potentially powerful for mapping extended line emission, but they also make it necessary to quantify the actual narrowband imaging performance, including image quality, photometric calibration, surface-brightness sensitivity, point-source detection limits, and field-dependent systematics.

In this paper, we present the initial performance tests of MASTA for wide-field narrowband imaging. \S~\ref{sect:obs} describes the telescope, the narrowband filter set, the QHY461 and COSMOS-8K cameras used for test observations, and the observations of M106 used in this work. We also present the data reduction and photometric procedures. \S~\ref{sect:results} characterizes the surface-brightness limits, point-source detection limits, the image quality, and the comparison between the two cameras. M106 and Cygnus Loop are further used as demonstration targets to test continuum subtraction and emission-line mapping with adjacent narrowband filters, followed by a quantitative comparison of the M106 H$\alpha$ surface brightness with an independent SITELLE measurement. \S~\ref{sect:conclusion} summarizes our major results and conclusions.

\section{Observations and Data Reduction}
\label{sect:obs}

\subsection{MASTA and its Camera Configurations}

MASTA is a wide-field optical facility with telescopes deployed at sites of Lenghu, Qinghai Province, China and Dayao, Yunnan Province, China. Operated by the Purple Mountain Observatory, it is a modular and scalable wide-field optical network optimized for high-efficiency sky surveys. The telescope is designed for diverse scientific goals, including the monitoring of transients (e.g. supernovae and gravitational wave counterparts)~\citep{2025AcASn..66...20Z}.

Each individual telescope has an aperture of 710 mm and a focal length of 1231 mm, resulting in a focal ratio of $f/1.7$. The prime focus configuration allows for a compact and high-throughput optical path. 

\begin{figure}[htbp]
\centering
\includegraphics[width=0.98\columnwidth]{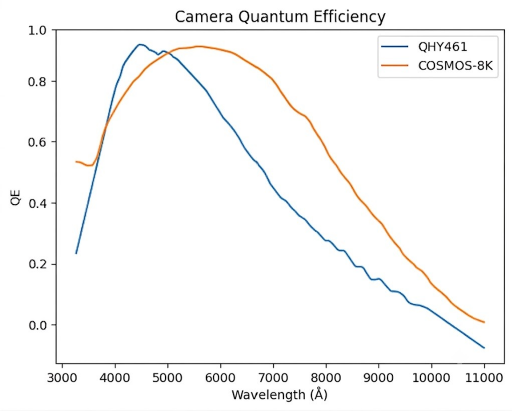}
\caption{
QE curves as a function of wavelength for the COSMOS-8K and QHY461 cameras.}
\label{fig: camera compare QE}
\end{figure}

The system can be equipped with two distinct cameras, COSMOS-8K and QHY461. Both cameras provide a wide FOV and low read noise. The COSMOS-8K provides a wider FOV of $3.75^\circ \times 3.75^\circ$, while QHY461 employs a smaller pixel scale ($0.626^{\prime \prime}\,\mathrm{pixel^{-1}}$). Both cameras use CMOS detectors. Detailed parameters for both sensors are summarized in Table~\ref{tab:camera_specs}. The quantum efficiency (QE) curves of the two cameras are shown in Figure~\ref{fig: camera compare QE}.

\begin{table}[htbp]
\begin{center}
\caption{Parameters of two cameras}
\label{tab:camera_specs}
\begin{tabular}{lcc}
\toprule
\textbf{Parameter} & \textbf{COSMOS-8K} & \textbf{QHY461} \\
\midrule
Pixel Array & $8120 \times 8120$ & $11760 \times 8842$ \\
Pixel Size & 10$\mu$m & 3.76$\mu$m \\
Pixel Scale & 1.662$''\,\mathrm{pixel}^{-1}$ & 0.626$''\,\mathrm{pixel}^{-1}$ \\
FOV & $3.75^\circ \times 3.75^\circ$ & $2.05^\circ \times 1.54^\circ$ \\
Read Noise & $< 1e^-$ & $1 \sim 3e^-$ \\
Dark Current & $\sim 0.1\,$ & $\sim 0.003$ (-20\,$^\circ$C) \\
Peak QE & $> 90\%$ & 91\% \\
\bottomrule
\end{tabular}
\end{center}

The pixel array refers to the total number of pixels listed for each camera, including the overscan region of QHY461. Pixel size refers to the physical dimensions of an individual pixel, while pixel scale represents the angle size of each pixel on the sky. Both cameras have low read noise, low dark current, and high peak quantum efficiency. Gain values are not listed because fixed, user-accessible gain values are not provided for the two cameras. The unit of dark current is $e^{-}\,\mathrm{pix}^{-1}\,\mathrm{s}^{-1}$. 
\end{table}

\subsection{Test Observations and Data Reduction}\label{subsec:reduction}

We obtained test observations over two nights targeting M106 (Messier 106, NGC 4258) and one night targeting the supernova remnant Cygnus Loop. The exposure times of two cameras are summarized in Table~\ref{tab:obs log}.

\begin{table*}[htbp]
\centering
\caption{Test Observations for M106 and Cygnus Loop.}
\label{tab:obs log}
\small
\begin{tabular}{llcccc}
\hline\noalign{\smallskip}
Target & Camera & Date & Filter ($\lambda_{\rm c}$) & Exposure & Total Exposure \\
\hline\noalign{\smallskip}
M106 & QHY461 & 2026-04-16 & 5170 \AA & $300\,\mathrm{s} \times 12$ & 3600\,s \\
M106 & QHY461 & 2026-04-16 & 6815 \AA & $300\,\mathrm{s} \times 12$ & 3600\,s \\
\noalign{\smallskip}
M106 & QHY461 & 2026-04-28 & 5005 \AA & $300\,\mathrm{s} \times 12$ & 3600\,s \\
M106 & QHY461 & 2026-04-28 & 5170 \AA & $300\,\mathrm{s} \times 12$ & 3600\,s \\
M106 & QHY461 & 2026-04-28 & 6544 \AA & $300\,\mathrm{s} \times 12$ & 3600\,s \\
M106 & QHY461 & 2026-04-28 & 6815 \AA & $300\,\mathrm{s} \times 12$ & 3600\,s \\
\hline\noalign{\smallskip}
Cygnus Loop & COSMOS-8K & 2025-08-30 & 5170 \AA & $30\,\mathrm{s} \times 20$ & 600\,s \\
Cygnus Loop & COSMOS-8K & 2025-08-30 & 6544 \AA & $30\,\mathrm{s} \times 20$ & 600\,s \\
\hline\noalign{\smallskip}
\end{tabular}
\end{table*}

\begin{figure}[htbp]
\centering
\includegraphics[width=0.98\columnwidth]{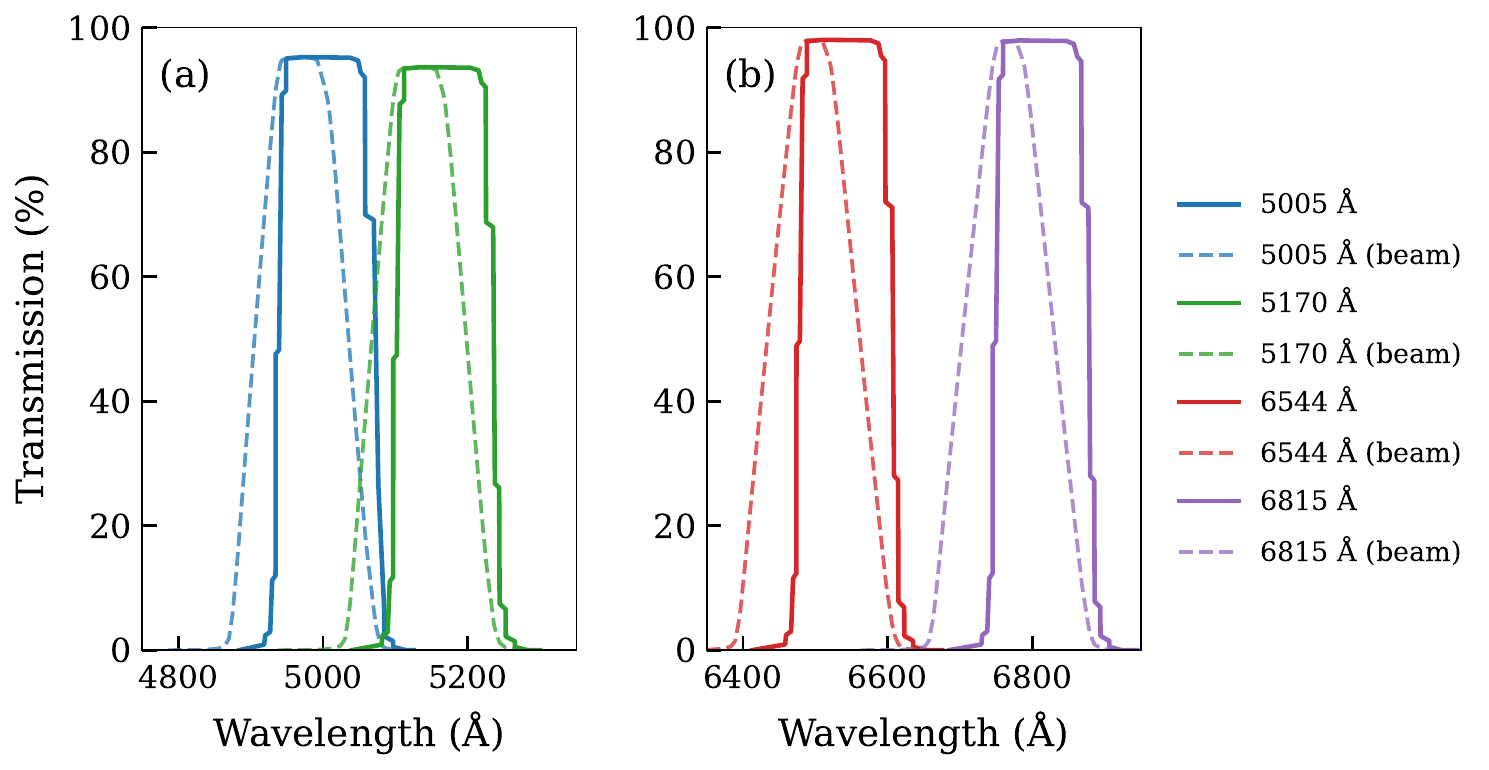}
\caption{
The transmission curves of four narrowband filters (centered on 5005\,\AA, 5170\,\AA, 6544\,\AA, and 6815\,\AA ) in the observations. The solid curves show the normal-incidence (parallel-beam) laboratory measurements, while dashed curves show the effective transmission of the filters integrated over the $f/1.7$ beam.} The beam-integrated profiles are shifted and broadened relative to the
parallel-beam curves. The central wavelengths are denoted in the legends, and detailed filter properties are listed in Table \ref{tab:filters}. These profiles only represent the intrinsic transmission of the filters.
\label{fig: filter transmission}
\end{figure}

Figure~\ref{fig: filter transmission} shows the transmission curves of four narrowband filters used in the observations, and their detailed specifications are listed in Table~\ref{tab:filters}. Both the figure and the table refer to laboratory measurements of the filters only and do not include the throughput of the optical system, detector, or atmosphere.

\begin{table*}[htbp]
\begin{center}
\caption{Narrowband Filters of the telescope}
\label{tab:filters}
\begin{tabular}{lcccccc}
\toprule
$\lambda_{\rm c}$ (\AA) & $W_{90}$ (\AA) & $W_{50}$ (\AA) & Peak Trans. (\%) & Thickness (mm) & Physical Size (mm) & Angular Size \\
\midrule
3745 & 153 & 193 & 94.53 & 7.99 & & \\
4860 & 168 & 216 & 93.72 & 7.98 & & \\
5005 & 151 & 191 & 95.26 & 7.98 & & \\
5170 & 152 & 203 & 93.65 & 8.00 & $117.91 \times 117.91$ & $5.49^\circ \times 5.49^\circ$ \\
6544 & 152 & 201 & 98.05 & 7.98 & & \\
6660 & 152 & 199 & 96.04 & 7.98 & & \\
6815 & 152 & 200 & 97.95 & 7.98 & & \\
9085 & 152 & 201 & 98.82 & 7.99 & & \\
\bottomrule
\end{tabular}
\end{center}

$\lambda_{\rm c}$ is the central wavelength of the filter. $W_{\rm 50}$ and $W_{\rm 90}$ are the full width at 50\% and 90\% peak transmission, respectively. Peak Transmission is the maximum transmission efficiency of the passband. 
\end{table*}

Given the focal ratio of $f/1.7$, the maximum off-axis angle reaches $\sim 18.48^\circ$, corresponding to a theoretical blueward shift up to $\sim 2.70\%$ ($\sim 177\,\mathrm{\mathring{A}}$ for the 6544\,\AA\ filter)~\citep{2025AcASn..66...58X}. Such a severe shift approaching the $W_{50}$ bandwidth can cause the targeted emission lines to drift near or beyond the bandpass edge. Large incidence angles and the finite beam cone also broaden and distort the transmission profile. The relatively broader passbands adopted for the narrowband filters, together with the use of adjacent narrowband filters as continuum references, help reduce the impact of these effects on continuum subtraction and wide-field imaging. Nevertheless, field-dependent variations in the effective bandpass may still introduce systematic uncertainties, particularly near the edge of the image. 

Data reduction is performed using the Python package \texttt{ccdproc}~\citep{ccdproc}, including bias correction, dark subtraction, flat-fielding for COSMOS-8K and overscan subtraction, flat-fielding for QHY461. The overscan subtraction of QHY461 is performed to remove the electronic baseline. Source detection and photometry are carried out using SExtractor~\citep{1996A&AS..117..393B}, while astrometric calibration is achieved using \texttt{astrometry.net}~\citep{2010AJ....139.1782L}. The available calibration frames need to be improved. For example, the dark frames exhibit localized non-uniformities (e.g. a small central bright region with a radius of approximately 150 pixels). The region is small compared to the image ($8120\times8120$ pixel for COSMOS-8K) and can be reduced by background subtraction. The calibration stars used for the zero point determination are distributed across the field, rather than being concentrated in this central region. Thus, it is unlikely that the region will introduce a significant systematic bias into the following flux calibration.

For astrometric alignment, we first extract a list of the 300 brightest, non-saturated, high-significance stellar coordinates using SExtractor and subsequently solve the World Coordinate System (WCS). Individual frames are then reprojected onto a common grid and aligned with the WCS solutions. To remove cosmic rays and prominent warm pixels, aligned images are initially median-combined in groups of three. These intermediate median-filtered frames are subsequently average-combined to produce the final deep co-added image, thereby maximizing the overall signal-to-noise ratio ($\mathrm{S/N}$).

SExtractor is applied to the stacked images, using \texttt{MAG\_AUTO} with Kron apertures to extract total fluxes and \texttt{MINIBACK\_RMS} to determine the local background root-mean-square (RMS) using a $512\times512$ pixel background mesh. The instrumental magnitude zero-points for our narrowband filters are calibrated using the Gaia DR3 spectra~\citep{2016A&A...595A...1G, 2023A&A...674A...1G}. We select non-variable stars within a Gaia G-band magnitude range of $12-16\,\mathrm{mag}$ and match them with SExtractor detected sources within a  radius of $2^{\prime\prime}$. Synthetic narrowband magnitudes are generated by convolving the Gaia spectra with the filter transmission curve (Figure~\ref{fig: filter transmission}) and the atmospheric transmission curve of Lenghu, Qinghai Province. We also test the inclusion of the camera QE curve in the convolving. The resulting change in the photometric zero point is $\le 0.001$ mag, indicating the QE has a negligible impact on the calibration for the present data. We use python package \texttt{Transmission Fitter}~\citep{2025A&A...699A..50G} to generate the atmospheric transmission curve. The atmospheric model is initialized with a typical local pressure of 615\,hPa for Lenghu and a precipitable water vapor of 4\,mm, corresponding to the measured value of April 2026. Other configurations are kept identical~\citep{2025A&A...699A..50G}. The photometric zero-point is then established by comparing these synthetic magnitudes with the corresponding instrumental count rates. The calibration stars selected from the same science frames approximately share the same atmospheric transparency and airmass. 

To assess the sensitivity of our photometry procedure to the finite spectral resolution of the Gaia XP spectra, we additionally convolve the spectra with Gaussian smoothing kernels corresponding to effective spectral resolutions of $R_{\rm smooth}=30-300$. The resulting zero-point shifts are small and converge with increasing spectral resolution. At the nominal Gaia XP resolution of $R_{\rm smooth}=100$, the zero-point shift is only $0.0026$ mag. The values are lower than the observed zero-point scatter of $0.0172$ mag, indicating that the spectral resolution of the Gaia XP spectra is not a dominant source of uncertainty in our flux calibration.

\section{Results}\label{sect:results}

\subsection{Surface Brightness Detection Limits}

Narrowband imaging provides an efficient way to identify line emission and map the spatial distribution of ionized gas. To characterize the sensitivity of the system for detecting extended emission, we calculate the surface brightness (SB) detection limits for each filter using a threshold of 3$\sigma$ relative to the background RMS.

\begin{figure}[htbp]
\centering
\begin{subfigure}[b]{0.49\textwidth}
    \centering
    \includegraphics[width=\textwidth]{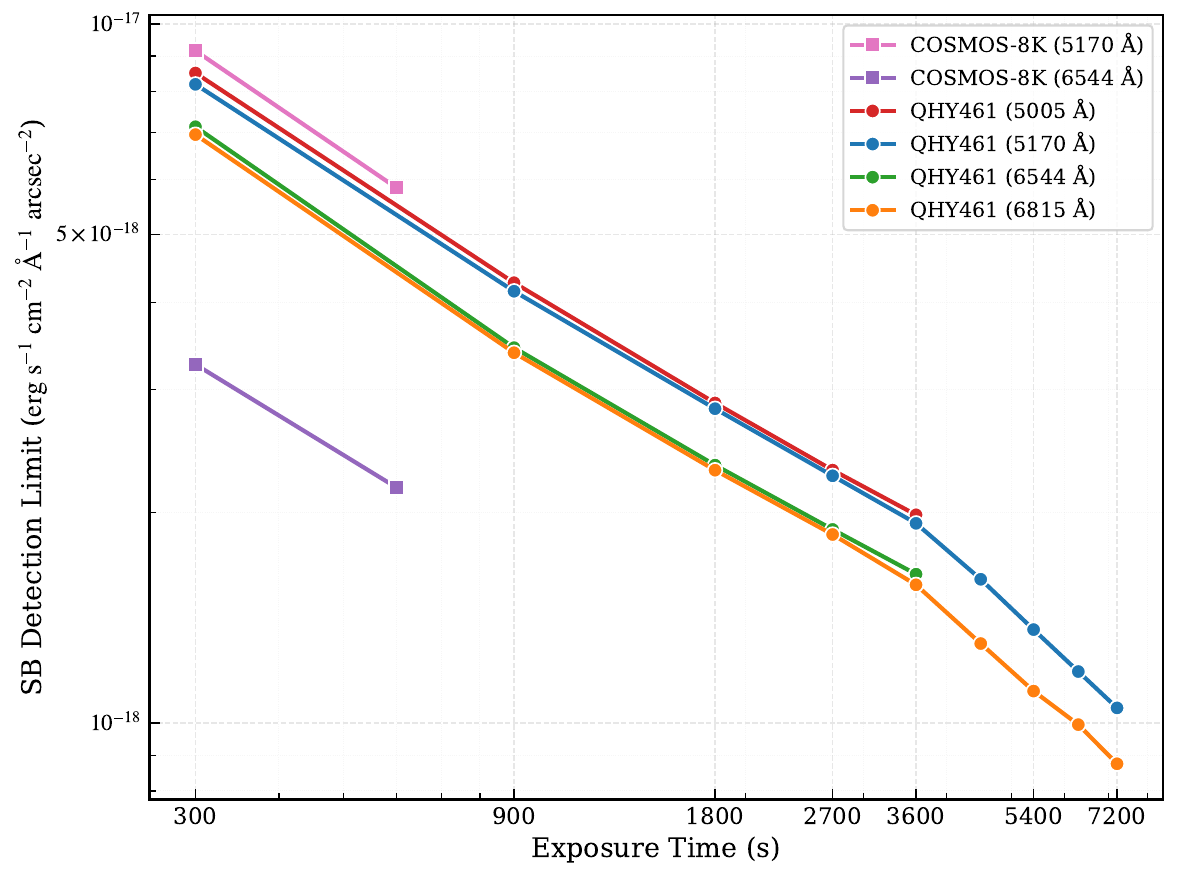}
    \caption{SB detection limits of the QHY461 and COSMOS-8K cameras as a function of exposure time for different filters in units of $\mathrm{erg\,s^{-1}\,cm^{-2}\,\mathrm{\mathring{A}}^{-1}\,arcsec^{-2}}$. Both axes are presented on a log scale. The dots represent the SB detection limit for QHY461 across four narrowband filters (5005\,\AA, 5170\,\AA, 6544\,\AA, and 6815\,\AA), while the squares represent COSMOS-8K with the 5170\,\AA\ and 6544\,\AA\ filters.}
    \label{fig:camera_compare_SB}
\end{subfigure}

\begin{subfigure}[b]{0.49\textwidth}
    \centering
    \includegraphics[width=\textwidth]{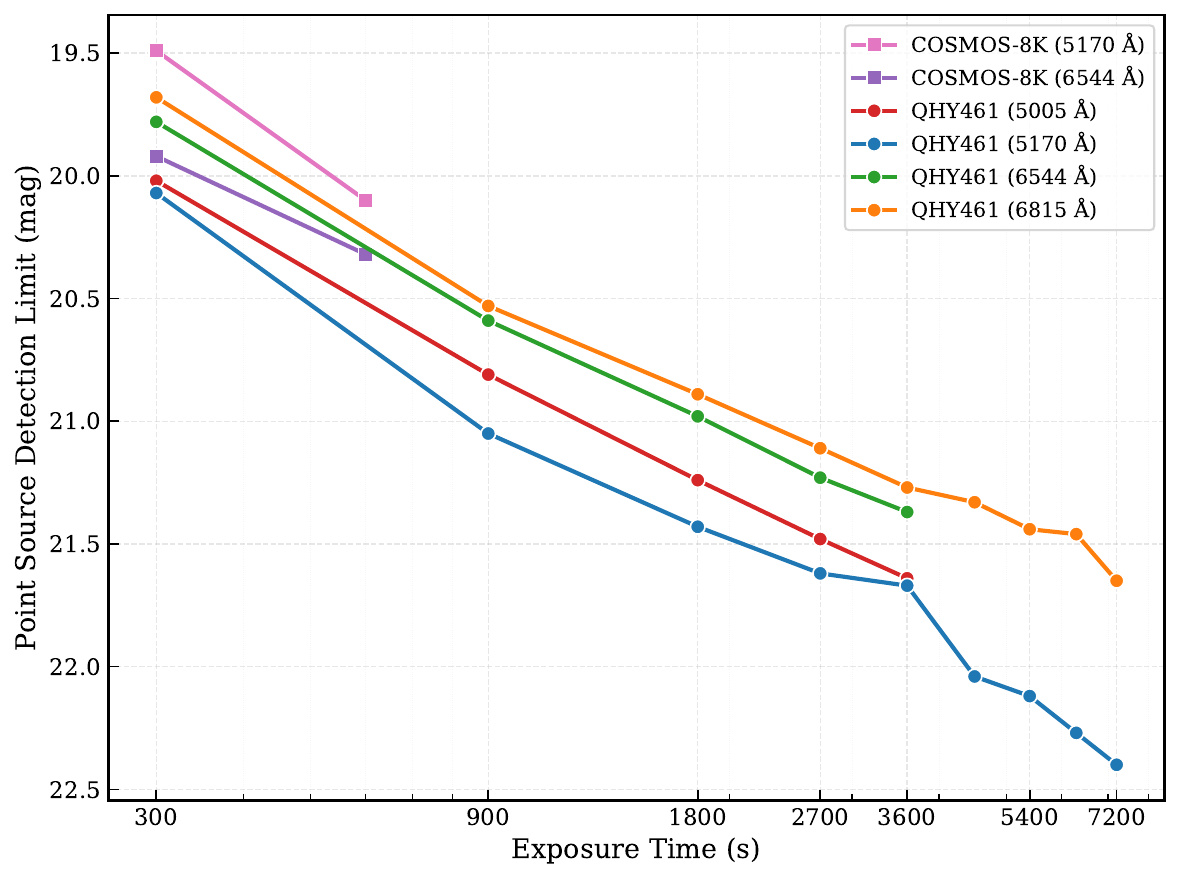}
    \caption{The $5\sigma$ point-source detection limits of the QHY461 and COSMOS-8K cameras as a function of exposure time for different filters in units of $\mathrm{mag}$. The limiting magnitude generally becomes deeper as the exposure time increases. Since the photometric aperture radius is defined as 1.5 times the image FWHM, variations in stellar FWHM lead to deviations from a simple exposure-time scaling, especially for the 5170\,\AA\ and 6815\,\AA\ filters.}
    \label{fig:camera_compare_source}
\end{subfigure}
\caption{Comparison of performance between the QHY461 and COSMOS-8K cameras: (a) SB detection limits and (b) $5\sigma$ point-source detection limits as a function of exposure time.}
\label{fig:camera_compare}
\end{figure}

Figure~\ref{fig:camera_compare_SB} illustrates the evolution of SB detection limits as a function of exposure time for the QHY461 and COSMOS-8K cameras. The results follow an approximately linear relation when plotted in magnitudes against $\log_{10}(t)$.

The COSMOS-8K generally reaches deeper SB detection limits than the QHY461, but the differences vary with wavelength. For example, with the 6544\,\AA\ filter, COSMOS-8K reaches $2.17 \times 10^{-18}\,\mathrm{erg\,s^{-1}}$ $\mathrm{cm^{-2}\,\mathrm{\mathring{A}}^{-1}\,arcsec^{-2}}$ at 600\,s, while QHY461 approaches $\sim 1.63 \times 10^{-18}\,\mathrm{erg\,s^{-1}\,cm^{-2}\,\mathrm{\mathring{A}}^{-1}\,arcsec^{-2}}$ with a much longer exposure of 3600\,s. The main reason is likely to be the higher quantum efficiency of COSMOS-8K at a wavelength of more than 6000\,\AA, as shown in Figure~\ref{fig: camera compare QE}. In contrast, with the bluer 5170\,\AA\ filter, the SB limit gap between the two cameras is smaller. In addition, the larger pixel scale of COSMOS-8K ($1.662^{\prime\prime}\,\mathrm{pixel^{-1}}$ for COSMOS-8K and $0.626^{\prime\prime}\,\mathrm{pixel^{-1}}$ for QHY461) may also reduce the number of pixels corresponding to a given angular area and therefore lower the impact of read noise. For COSMOS-8K, the SB limit shows an approximately linear improvement with $\log_{10}(t)$ with a slope of \ 
$\sim -0.6$. These results suggest that COSMOS-8K could be advantageous for mapping diffuse low-surface-brightness emission. QHY461 provides finer image sampling with its smaller pixel scale.

To provide a guide for prospective observational planning with the QHY461, we parameterize the relation between the SB detection limit and exposure time as 

\begin{equation}
\label{eq:SBlimit_flux}
\log_{10} \left( F_{\mathrm{limit}} \right) = C \log_{10} \left( \frac{t}{3600\,\mathrm{s}} \right) + D
\end{equation}

where $F_\mathrm{{limit}}$ is the SB limit, $t$ is the total exposure time in seconds, $C$ is the slope reflecting the scaling efficiency ($F_{\mathrm{limit}} \propto t^C$), and $D$ is the constant for normalization. In the ideal background-limited regime, the noise is dominated by Poisson fluctuations of the sky background. For a fixed detection threshold, the SB limit scales as $t^{-0.5}$, corresponding to a slope of $C = -0.5$. The best-fit parameters are shown in Table~\ref{tab:fit results}. The fitted slopes range from $-0.63$ to $-0.59$, which are steeper than the theoretical value of $-0.5$. This can arise from the variation in observing conditions (e.g., transparency or lunar phase) across different exposures.

\begin{table}[htbp]
\centering
\caption{Best-fit Parameters of Four Filters for the QHY461 Camera}
\label{tab:fit results}
\begin{tabular}{lccc}
\toprule
Filter (\AA) & $C$ & $D$ \\
\midrule
5005 & $-0.59 \pm 0.01$ & $-17.71 \pm 0.01$ \\
5170 & $-0.59 \pm 0.02$ & $-17.75 \pm 0.01$ \\
6544 & $-0.63 \pm 0.02$ & $-17.80 \pm 0.01$ \\
6815 & $-0.59 \pm 0.02$ & $-17.84 \pm 0.01$ \\
\bottomrule
\end{tabular}

\begin{flushleft}
The table lists the fitted values of $C$ and $D$ and their uncertainties, the parameters of Equation~\ref{eq:SBlimit_flux}.
\end{flushleft}
\end{table}

\subsection{FWHM and Point-Source Detection Limit} \label{sec: FWHM and limit}

We measure the full width at half maximum (FWHM) of bright, isolated stars with $\mathrm{S/N}>10$. Stellar FWHM is measured with SExtractor, and the median value is adopted as the representative FWHM for each filter. Table~\ref{tab:fwhm} summarizes the median FWHM values, spatial variations and sampling rates across different filters.

\begin{table*}[htbp]
\begin{center}
\caption{FWHM Statistics and Sampling Properties for the Two Cameras.}
\label{tab:fwhm}
\small 
\begin{tabular}{llcccc}
\toprule
Camera & Pixel Scale ($^{\prime\prime}$\,pixel$^{-1}$) & Filter ($\lambda_{\rm c}$) & FWHM ($^{\prime\prime}$) & Field Variation ($^{\prime\prime}$) & Sampling (pixels) \\
\midrule
COSMOS-8K & 1.662 & 5170 \AA & 2.89 & 0.08 & 1.7 \\
COSMOS-8K & 1.662 & 6544 \AA & 3.95 & 0.12 & 2.4 \\
\addlinespace 
\hline
\addlinespace
QHY461    & 0.626 & 5005 \AA & 2.18 & 0.009 & 3.5 \\
QHY461    & 0.626 & 5170 \AA & 1.87 & 0.09  & 3.0 \\
QHY461    & 0.626 & 6544 \AA & 2.00 & 0.15  & 3.2 \\
QHY461    & 0.626 & 6815 \AA & 2.08 & 0.17  & 3.3 \\
\bottomrule
\end{tabular}
\end{center}

Field variation is defined as the RMS of a two-dimensional quadratic polynomial fit ($\mathrm{FWHM} = a + bx + cy + dx^2 + ey^2 + fxy$) to the FWHM distribution across the FOV, indicating the spatial inhomogeneity of the PSF. The PSF sampling rate is defined as $\mathrm{FWHM / pixel\ scale}$. The values are measured with a total exposure time of 3600\,s for the QHY461 camera and 600\,s for the COSMOS-8K camera.
\end{table*}

The COSMOS-8K image with 5170\,\AA\ filter shows a low sampling rate of $\sim 1.7$ in Table~\ref{tab:fwhm}. To assess the impact of under-sampling, we perform Monte Carlo simulations using Gaussian and Moffat PSFs at the COSMOS-8K and QHY461 pixel scales. The coarse sampling introduces a systematic FWHM bias of approximately $0.16^{\prime\prime}$--$0.26^{\prime\prime}$ when pixel integration is not explicitly modeled, whereas the pure sub-pixel phase scatter is $\lesssim0.014^{\prime\prime}$. For aperture photometry with a radius of $1.5\times$FWHM and a fixed sky surface brightness, the resulting flux scatter is $1.1\%$--$1.4\%$, comparable to that obtained for QHY461. Therefore, the sampling-induced bias is modest compared with the intrinsic FWHM and the field-to-field variation in image quality. 

To estimate the magnitude limits for point sources, we adopt circular apertures with a radius of 1.5 times the stellar FWHM. The local background RMS estimated from the corresponding background mesh is used to determine the corresponding 5$\sigma$ limiting magnitudes. At fixed sky surface brightness, the aperture-flux scatter with a 1.5-FWHM aperture is $1.1–1.4\%$, comparable to that obtained for QHY461.

Figure~\ref{fig:camera_compare_source} shows the 5$\sigma$ point-source detection limits as a function of the exposure time for two cameras. The limiting magnitudes generally become deeper with increasing exposure time; however, variations in stellar FWHM introduce scatter and offsets among different filters relative to a simple theoretical scaling.

\begin{figure}[htbp]
\centering
\includegraphics[width=0.98\columnwidth]{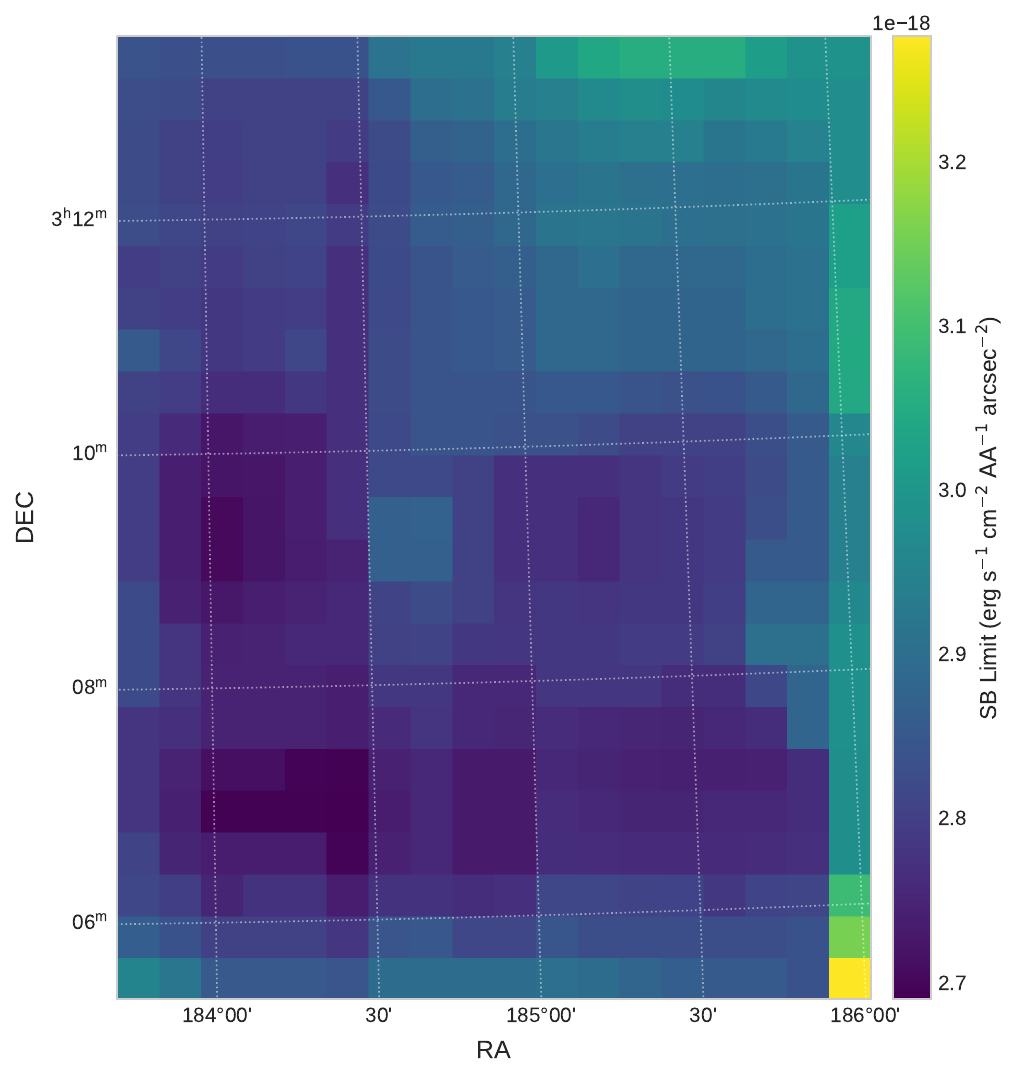}
\caption{
Two-dimensional distribution of the $3\sigma$ SB detection limit across the full FOV of the QHY461 camera. The map is derived from a representative 3600\,s exposure using the 6544\,\AA\ narrowband filter. Each grid in this figure represents a background mesh size of 512 pixels, while the local background RMS is converted into $3\sigma$ SB limits in units of $\mathrm{erg\,s^{-1}\,cm^{-2}\,\mathrm{\mathring{A}}^{-1}\,arcsec^{-2}}$. }
\label{fig: SB limit map}
\end{figure}

Variations in the detector response and optical system aberrations can lead to non-uniform backgrounds across the FOV. Figure~\ref{fig: SB limit map} shows the spatial variation of the SB detection limit.

\subsection{Continuum-Subtracted \texorpdfstring{H$\alpha$}{Hα} and [O\,III] Images}

The H$\alpha$ emission is widely used to trace the distribution of ionized gas in galaxies~\citep{1993A&A...268..419C}. Using four combined narrowband images, we perform continuum subtraction to map H$\alpha$ and [\ion{O}{III}] emission in M106. The filters centered on 6544\,\AA \ and 5005\,\AA \ are used as the emission-line frames, while the 6815\,\AA \ and 5170\,\AA \ filters are adopted as corresponding continuum references.

Background subtraction is performed using the SExtractor \texttt{BACKGROUND} model with a mesh size of 512 pixels to minimize oversubtraction of extended emission. All images are reprojected onto a common WCS grid prior to continuum subtraction. Initial cross-band scaling factors are derived from the median flux ratios of non-saturated, isolated stars cross-matched between the emission-line and continuum frames. 

\begin{figure*}[htbp]
\centering
\begin{minipage}[b]{0.49\textwidth}
\centering
\includegraphics[width=\textwidth]{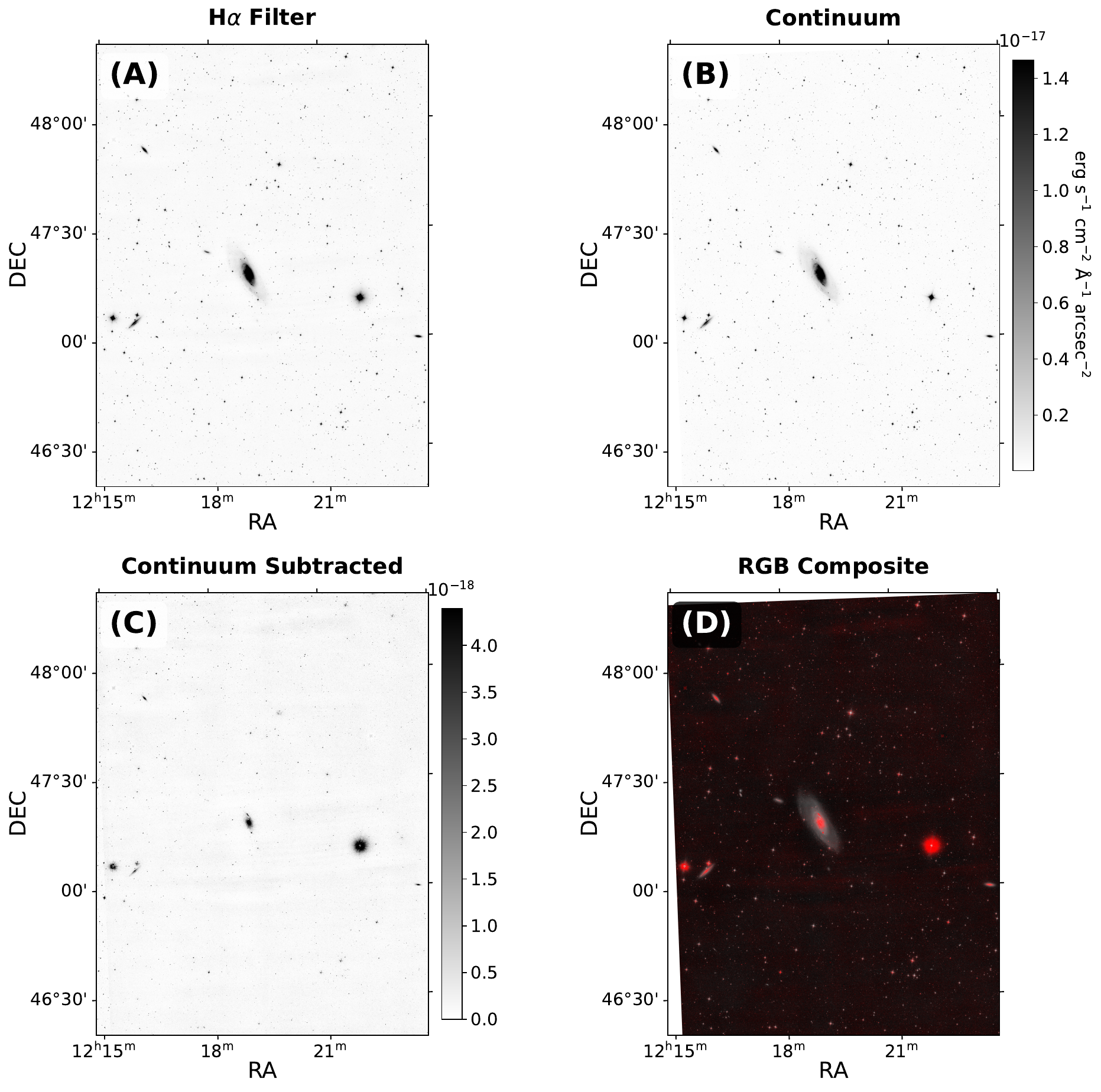}
\end{minipage}
\hfill
\begin{minipage}[b]{0.49\textwidth}
\centering
\includegraphics[width=\textwidth]{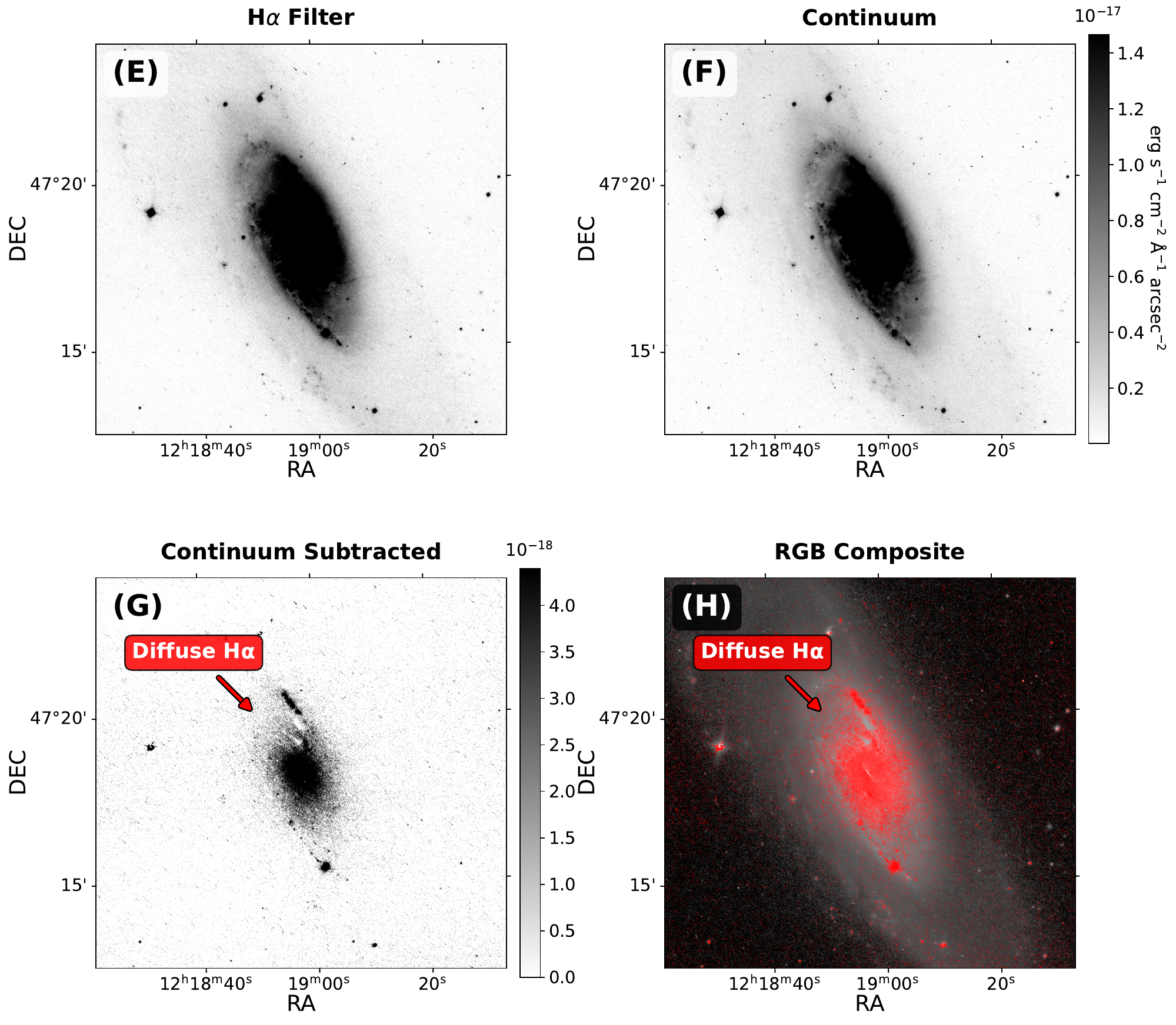}
\end{minipage}

\caption{H$\alpha$ emission line distribution and zoomed-in details in M106. 
\textbf{Left panels:} Full example M106 images taken with the QHY461 camera. (A) Narrowband image obtained with the 6544\,\AA\ filter covering the H$\alpha$ emission and stellar continuum; (B) continuum reference image constructed from the 6815\,\AA\ and 5170\,\AA \ filter; (C) continuum-subtracted H$\alpha$ image of M106; (D) RGB color-composite image, in which the continuum image is included equally in the red, green, and blue channels, while the continuum-subtracted 6544\,\AA\ image is added to the red channel. The continuum therefore appears approximately white, whereas regions with excess H$\alpha$ emission appear red. A linear stretch is applied to the continuum-subtracted H$\alpha$ emission to enhance the visibility of the emission regions. Panels (A), (B), and (C) are calibrated to a common flux density unit of $\mathrm{erg\,s^{-1}\,cm^{-2}\,\mathrm{\mathring{A}}^{-1}\,arcsec^{-2}}$.
\textbf{Right panels:} Zoomed-in view of the central region surrounding M106. Panels (E)-(G) show the zoomed-in versions of Panels (A)-(C), respectively. Panel (H) shows the detailed color-composite map revealing the emission regions in the disk and arms, using the same channel assignment and scaling as Panel (D). The extended H$\alpha$ emission is highlighted by the ``Diffuse H$\alpha$'' label in Panels (G) and (H).}
\label{fig:Halpha_combined}
\end{figure*}

\begin{figure*}[htbp]
\centering
\begin{minipage}[b]{0.49\textwidth}
\centering
\includegraphics[width=\textwidth]{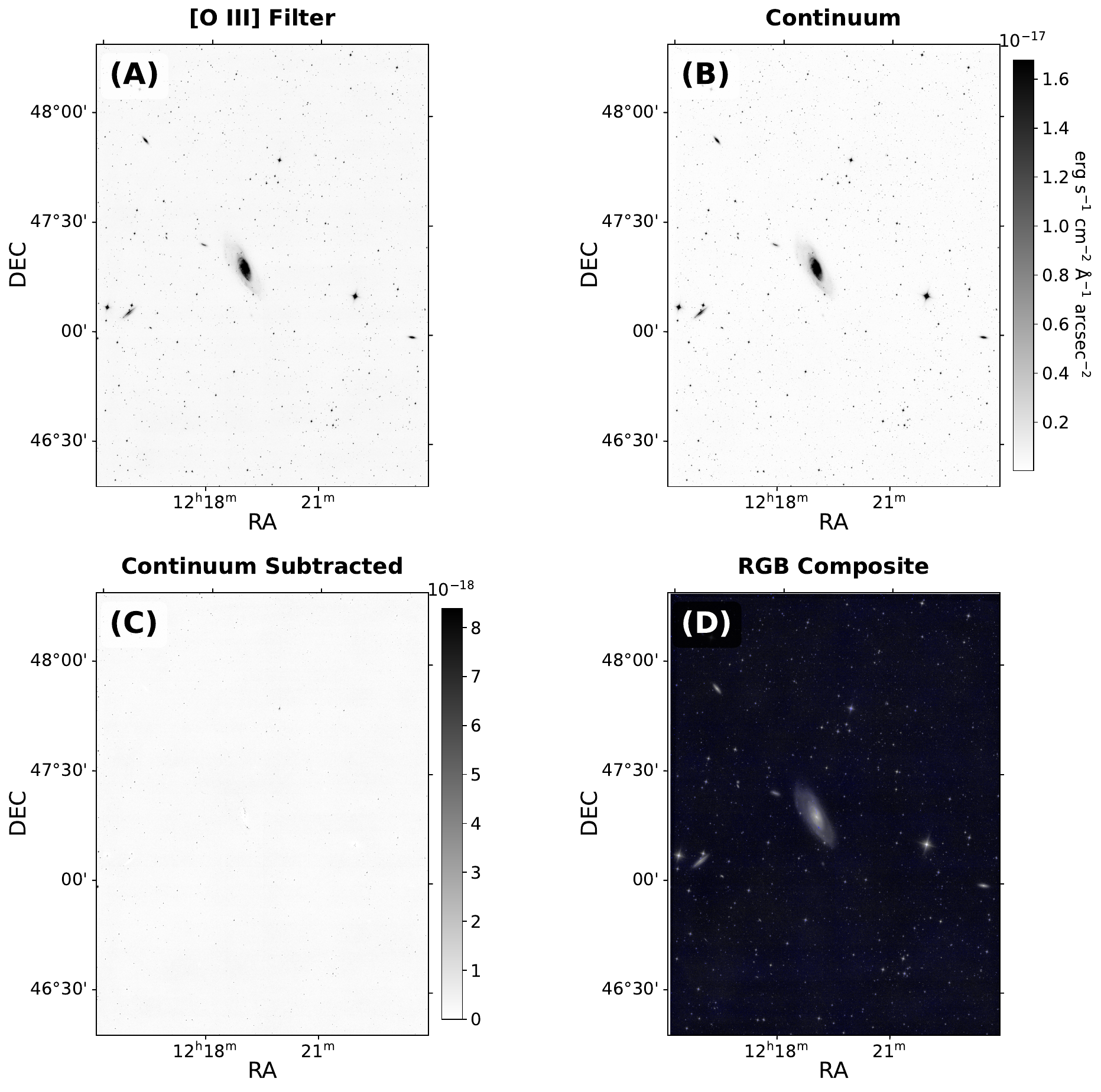}
\end{minipage}
\hfill
\begin{minipage}[b]{0.49\textwidth}
\centering
\includegraphics[width=\textwidth]{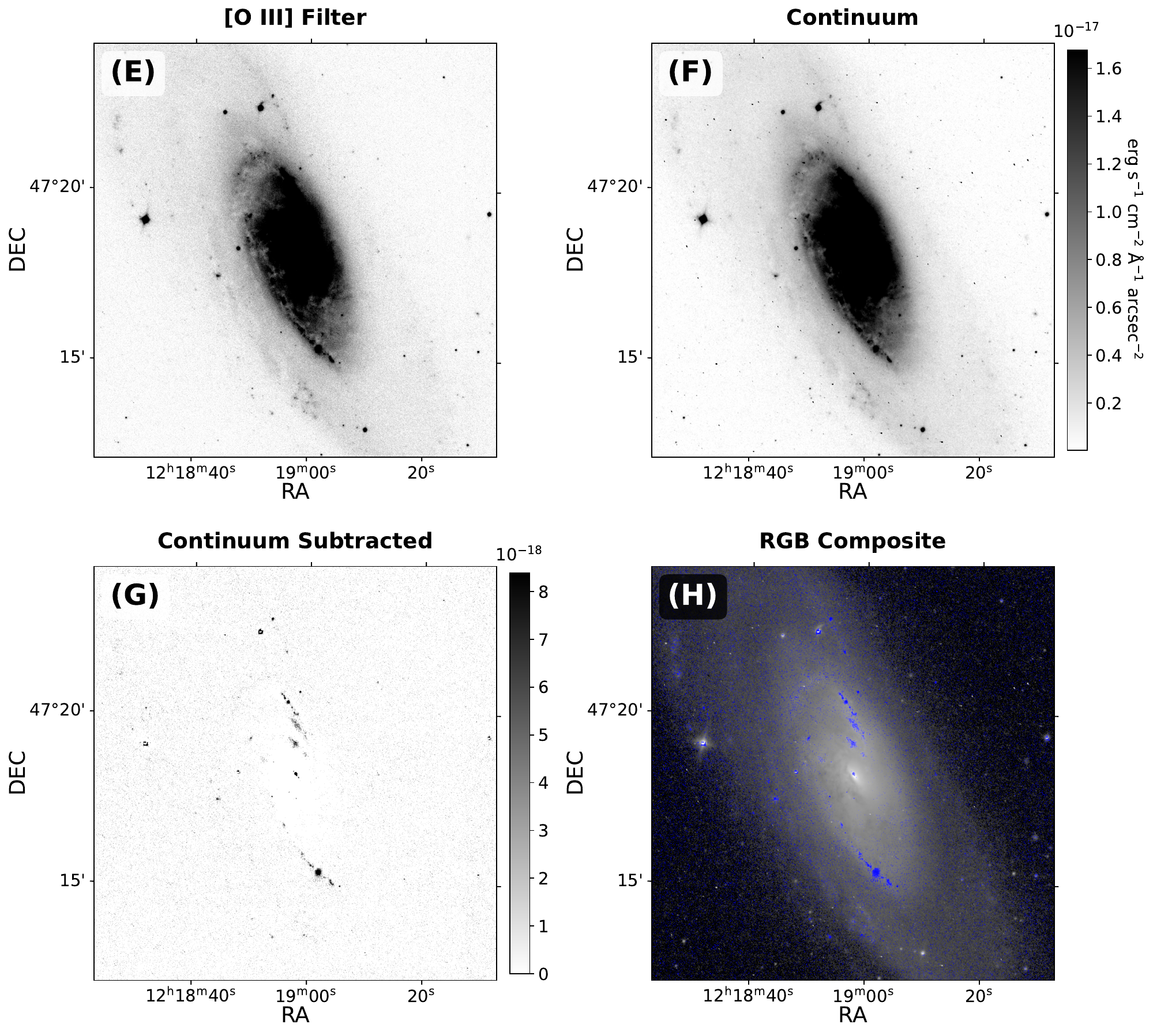}
\end{minipage}

\caption{[\ion{O}{III}] emission and zoomed-in details in M106, same as Fig.~\ref{fig:Halpha_combined}. 
\textbf{Left panels:} Full images, where the 5005\,\AA\ filter is used as the emission-line frame and a combined continuum frame is derived using the 5170\,\AA\ and 6815\,\AA\ filters. Panel (D) is an RGB color composite, in which the continuum image is included equally in the red, green, and blue channels, while the continuum-subtracted 5005,\AA\ image is stretched and added to the blue channel. The continuum therefore appears approximately white, whereas regions with excess [O III] emission appear blue.
\textbf{Right panels:} Zoomed-in details of the [\ion{O}{III}] emission.}
\label{fig:OIII_combined}
\end{figure*}

To minimize the effect of inter-band PSF variations, we perform PSF matching of all frames to the largest FWHM among them using Gaussian convolution kernels. The calibrated continuum images are then aligned to the emission line frame, scaled by their stellar flux ratios, and median-combined to construct the final continuum reference. However, this procedure assumes a spatially invariant PSF and does not account for the detailed PSF shape or its spatial variation across the wide field. Residuals around bright stars remain visible in the continuum-subtracted images after initial PSF matching. These are likely caused by non-Gaussian PSF structure, spatial PSF variations, small astrometric offsets, focus variations, and uncertainties in the scaling factor.

Panels (A)-(D) of Figure~\ref{fig:Halpha_combined} present the continuum-subtracted H$\alpha$ image of the full FOV, while Panels (E)-(H) detail the zoom-in image of M106. The $3\sigma$ SB limit is approximately $1.63 \times 10^{-18}\,\mathrm{erg\,s^{-1}\,cm^{-2}\, \mathring{A}^{-1}\,\mathrm{arcsec^{-2}}}$ (for the 6544\,\AA \ filter with a total exposure time of 3600\,s). H$\alpha$ emission is mainly detected in the central region of the galaxy. A faint stripe pattern with an amplitude of $\sim 4$ counts, comparable to the background RMS, is present in the H$\alpha$ image and continuum-subtracted image. Owing to its low amplitude, the feature is not readily visible in the displayed images. The feature is also present in contemporaneous white-light (unfiltered) and 6544\,\AA\ flat-field frames, but is absent in the bias frames, suggesting that it is more likely to arise from the optical system rather than the detector-related effects. These features can guide future baffling and stray-light control, while in array mode, a unit with abnormal optical performance can be excluded from science observations. The corresponding results for [\ion{O}{III}] are presented in Figure~\ref{fig:OIII_combined}. The continuum-subtracted image does not show evidence for a systematic negative residual around the central region of M106. Compared with the widespread H$\alpha$ emission across the galactic disk and spiral arms of M106, the [\ion{O}{III}] emission is considerably weaker and confined to smaller regions. The continuum-subtracted images reveal diffuse H$\alpha$ emission across the disk and spiral arms, demonstrating the capability of mapping diffuse ionized gas. 

\begin{figure*}[t]
\centering
\includegraphics[width=0.49\textwidth]{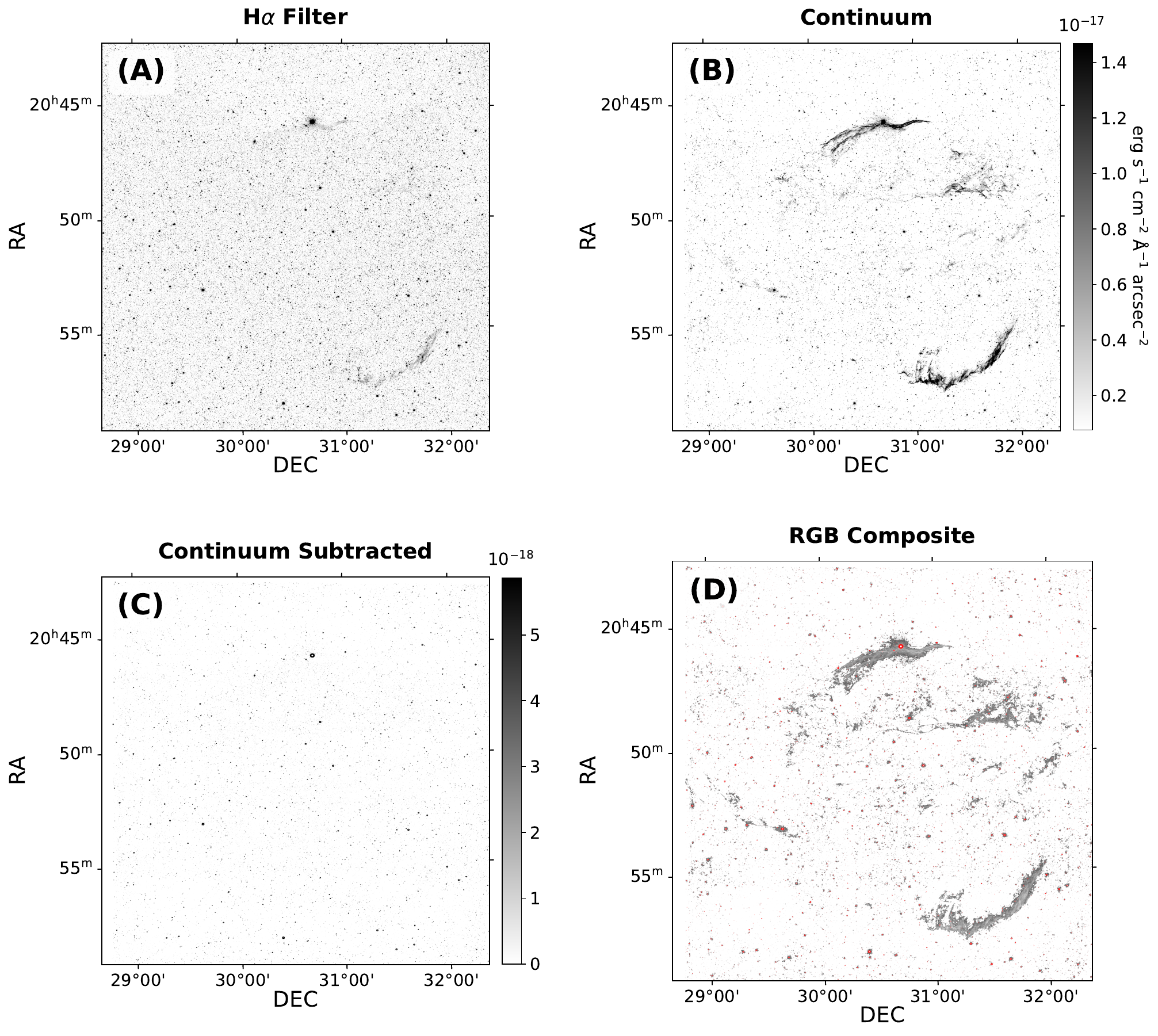}
\hfill
\centering
\includegraphics[width=0.49\textwidth]{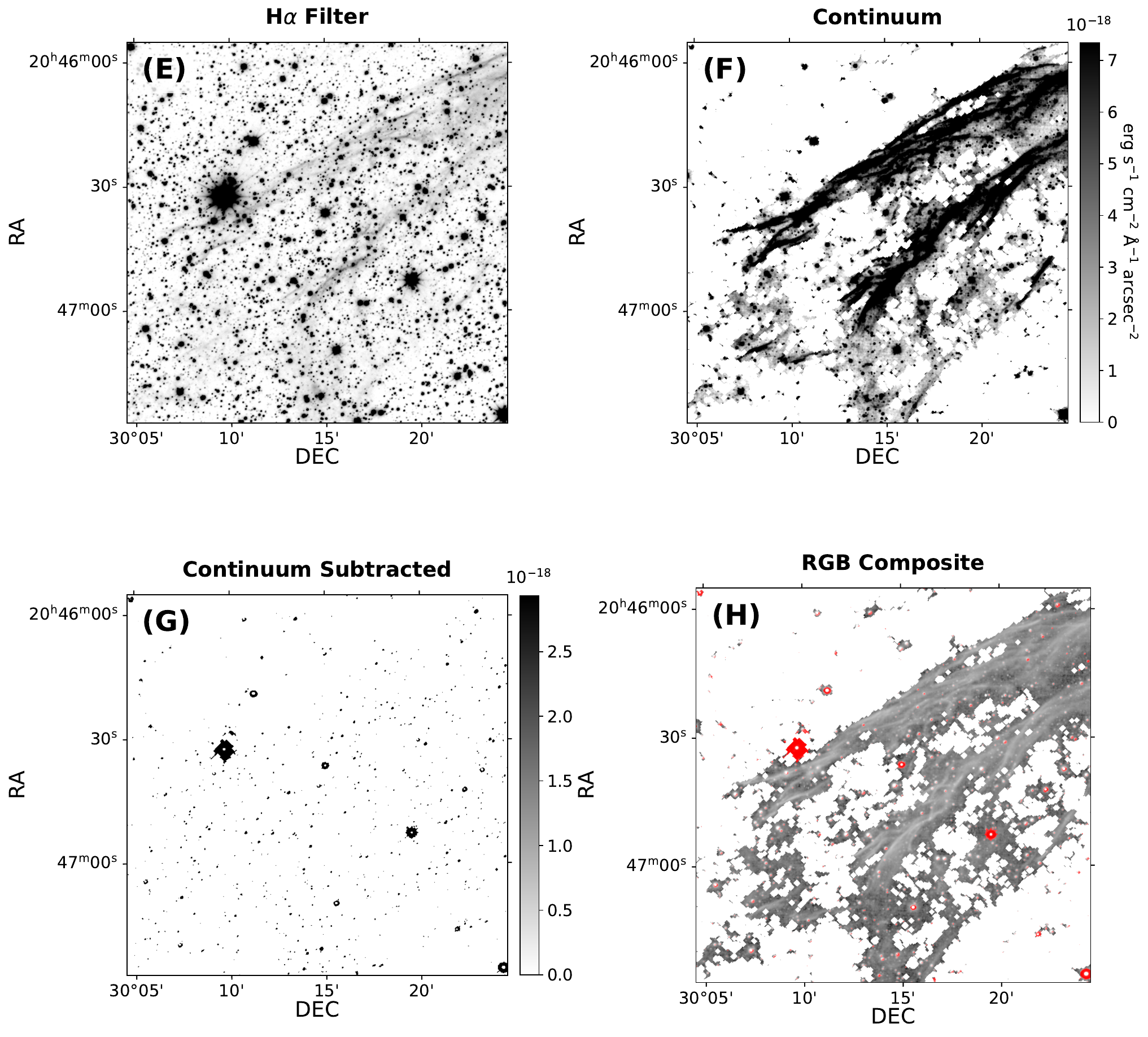}

\caption{Continuum-subtracted H$\alpha$ imaging of the Cygnus Loop with the COSMOS-8K camera. 
\textbf{Left panels:} Full image targeting the Cygnus Loop. The 6544\,\AA\ filter is used as the emission-line frame in Panel (A) and a combined continuum baseline is derived using the 5170\,\AA\ filter. Panel (C) shows the continuum-subtracted image, and Panel (D) shows the color-composite image with the net H$\alpha$ line emission explicitly color-coded into the red channel. 
\textbf{Right panels:} Zoom-in images of filaments of the Cygnus Loop. Clear structure filamentary is visible in Panel (F) with the 5170\,\AA\ filter. The ring-like residuals in Panel (G) arise from image quality, residual differences in PSF shape and spatial variation between the observations with the 6544\,\AA\ and 5170\,\AA\ filters.}
\label{fig:Cygnus_Loop_combined}
\end{figure*}
\par

The Cygnus Loop exhibits markedly different appearances in the 6544\,\AA\ and 5170\,\AA\ images. The 6544\,\AA\ frame exhibits broader stellar profiles and less distinct filamentary structures, suggesting poorer image quality, likely due to less favorable atmospheric conditions. In contrast, the 5170\,\AA\ image reveals sharp and well-defined filaments across the nebula.

The COSMOS-8K sampling is unlikely to be the dominant cause of these residuals. As discussed in \S~\ref{sec: FWHM and limit}, Monte Carlo simulations show that the sampling-induced FWHM bias is substantially smaller than the observed inter-band FWHM difference of $\sim1.06^{\prime\prime}$. The dominant limitation for continuum subtraction is therefore more likely to be the different image quality, rather than the under-sampling of the COSMOS-8K detector.

 Although the differing image quality prevents a direct comparison between the two bands and contributes to the stellar residuals visible in the continuum-subtracted image, the clear visibility of the extended gas structures in the 5170\,\AA\ image demonstrates the ability of the system to detect low-surface-brightness extended structures across a wide field.

The present continuum-subtracted images are intended primarily as demonstrations of the narrowband imaging capability during commissioning. Because the available calibration frames are not yet optimal, the continuum scaling is empirically adjusted, and the effective filter bandpass may vary across the wide field; we do not interpret the line maps as precise line-flux measurements. Future work incorporating improved calibration frames, spatially varying PSF models and position-dependent throughput corrections will further improve the accuracy of quantitative diffuse-emission measurements. 

\subsection{Comparison of M106 \texorpdfstring{H$\alpha$}{Hα} Map} \label{sec: comparsion of m106}

\begin{figure*}[htbp]
\centering
\includegraphics[width=0.98\textwidth]{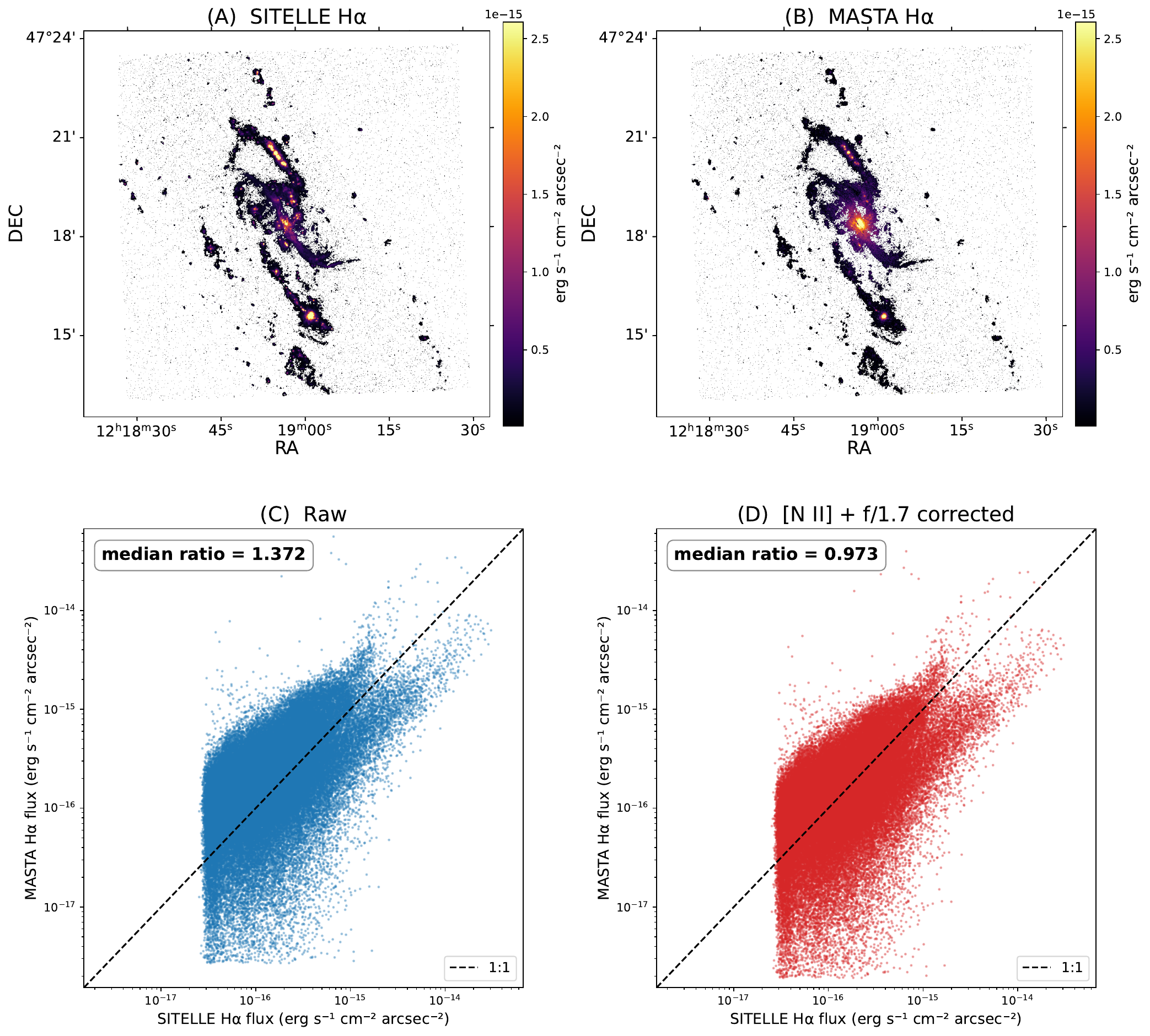}
\caption{
Panel (A) shows the surface brightness H$\alpha$ map from \citealt{2025MNRAS.542...25F}, while Panel (B) is the continuum-subtracted H$\alpha$ image of MASTA. The two panels use the same colorbar range, and the spatial mask applied to the SITELLE map is also applied to the MASTA image for a direct visual comparison. The diffuse H$\alpha$ emission highlighted in the right panels of Figure~\ref{fig:Halpha_combined} appears partly absent here because the SITELLE map masks spaxels with H$\alpha$ S/N $<3$, thereby excluding part of the faint, extended emission from both panels. Panel (C) shows the pixel-by-pixel comparison between the SITELLE and MASTA H$\alpha$ surface brightnesses before correction, while Panel (D) shows the result after accounting for [\ion{N}{II}] contamination and the wavelength-dependent filter response in the $f/1.7$ converging beam.}
\label{fig: m106 compare}
\end{figure*}

To quantitatively assess the scientific capability and photometric calibration of MASTA, we compare the continuum-subtracted image of M106 with the surface brightness H$\alpha$ map obtained with SITELLE by \cite{2025MNRAS.542...25F}. The SITELLE map is provided in logarithmic surface-brightness units of \(\log_{10}[{\rm erg\,s^{-1}\,cm^{-2}\,arcsec^{-2}}]\), and the spaxels with H$\alpha$ $\mathrm{S/N<3}$ are masked. For a consistent spatial comparison, we apply the same spatial mask to the MASTA image. Consequently, the faint, extended H$\alpha$ emission highlighted in the right panels of Figure~\ref{fig:Halpha_combined} is also partly excluded from Figure~\ref{fig: m106 compare}. For the pixel-by-pixel flux comparison, we additionally require the MASTA H$\alpha$ S/N to be greater than 3.

As shown in Figure~\ref{fig: filter transmission}, the 6544\,\AA\ filter of MASTA can also cover the [\ion{N}{II}] emission lines. We therefore estimate the [N II] contribution of [\ion{N}{II}] by adopting a representative $([\mathrm{N\,II}]\lambda6583/\mathrm{H}\alpha)$ ratio of 0.5\cite{2010MNRAS.406..181J} and a theoretical $[\mathrm{N\,II}]\lambda6583/\lambda6548$ ratio of 3. The wavelength-dependent shift of the transmission curve of the 6544\,\AA\ filter in the $f/1.7$ converging beam is also taken into account. We finally obtain a median ratio of 1.372 and a corrected ratio of 0.973 (Panel~D of Figure~\ref{fig: m106 compare}), demonstrating good agreement between the MASTA narrowband photometry and the independent SITELLE measurement.

\section{Conclusion}
\label{sect:conclusion}

In this paper, we present initial performance tests of MASTA for wide-field narrowband imaging and establish a data reduction and photometry pipeline. The measured surface-brightness limits become deeper with increasing exposure time and can be described by empirical relations useful for planning future observations, although the fitted slopes should be interpreted in the context of varying observing conditions and residual calibration uncertainties.

The two camera configurations provide complementary capabilities. For the current commissioning data, the COSMOS-8K generally reaches deeper surface-brightness limits, especially in the red filters, while QHY461 provides finer image sampling and smaller measured stellar FWHMs. Continuum-subtracted images of M106 reveal H$\alpha$ emission associated with the galactic disk, and COSMOS-8K observations of the Cygnus Loop reveal extended filamentary emission. The comparison between the MASTA
and SITELLE surface brightness H$\alpha$ maps of M106 shows good consistency of photometry after accounting for [\ion{N}{II}] contamination and the wavelength-dependent filter response in the $f/1.7$ beam.

These results demonstrate the potential of MASTA for wide-field narrowband surveys and studies of diffuse ionized gas. Further studies will require improved calibration data, corrections for field-dependent throughput variations, and more accurate spatially varying PSF modeling and matching.

\section*{acknowledgements}

We are grateful to Luan Luan, Xiao Xu, and Chao Yang for their help editing the paper. We also thank D. Fernández-Arenas for providing the processed SITELLE surface brightness H$\alpha$ map of M106 (NGC 4258) used in the quantitative comparison presented in this work.

We also thank ChatGPT (OpenAI), Gemini (3.5, Google) and DeepSeek for language and grammar checks within the article. We carefully reviewed, edited, and revised the AI-generated texts to our own preferences, assuming ultimate responsibility for the content of the publication.

Jiang-Tao Li acknowledges the financial support of the National Natural Science Foundation of China (NSFC) through grants 12273111 and 12321003, from the China Manned Space Program through grants CMS-CSST-2025-A10 and CMS-CSST-2025-A04, and the Jiangsu Innovation and Entrepreneurship Talent Team Program through grant JSSCTD202436. 

\section*{Author Contributions}
Mingxuan Xu implemented the study and wrote the paper. Jiang-Tao Li conceived the idea, and Chen Zhang provided the data. All authors read and approved the final manuscript.
 
\section*{Declaration of Interests}
The authors declare no competing interests.



\bibliographystyle{ati} 
\bibliography{ati}      

@ARTICLE{2012ARA&A..50..531K,
       author = {{Kennicutt}, Robert C. and {Evans}, Neal J.},
        title = "{Star Formation in the Milky Way and Nearby Galaxies}",
      journal = {\araa},
         year = 2012,
        month = sep,
       volume = {50},
        pages = {531-608},
          doi = {10.1146/annurev-astro-081811-125610},
archivePrefix = {arXiv},
       eprint = {1204.3552},
 primaryClass = {astro-ph.GA},
       adsurl = {https://ui.adsabs.harvard.edu/abs/2012ARA&A..50..531K}
}

@ARTICLE{2019ApJ...881...26V,
       author = {{Vargas}, Carlos J. and {Walterbos}, Ren{\'e} A.~M. and {Rand}, Richard J. and {Stil}, Jeroen and {Krause}, Marita and {Li}, Jiang-Tao and {Irwin}, Judith and {Dettmar}, Ralf-J{\"u}rgen},
        title = "{CHANG-ES. XVII. H{\ensuremath{\alpha}} Imaging of Nearby Edge-on Galaxies, New SFRs, and an Extreme Star Formation Region{\textemdash}Data Release 2}",
      journal = {\apj},
         year = 2019,
        month = aug,
       volume = {881},
       number = {1},
          eid = {26},
        pages = {26},
          doi = {10.3847/1538-4357/ab27cb},
archivePrefix = {arXiv},
       eprint = {1906.07763},
 primaryClass = {astro-ph.GA},
       adsurl = {https://ui.adsabs.harvard.edu/abs/2019ApJ...881...26V}
}

@ARTICLE{2023MNRAS.519.6098L,
       author = {{Lu}, Li-Yuan and {Li}, Jiang-Tao and {Vargas}, Carlos J. and {Beck}, Rainer and {Bregman}, Joel N. and {Dettmar}, Ralf-J{\"u}rgen and {English}, Jayanne and {Fang}, Taotao and {Heald}, George H. and {Li}, Hui and {Qu}, Zhijie and {Rand}, Richard J. and {Stein}, Michael and {Wang}, Q. Daniel and {Wang}, Jing and {Wiegert}, Theresa and {Zheng}, Yun},
        title = "{eDIG-CHANGES I: extended H{\ensuremath{\alpha}} emission from the extraplanar diffuse ionized gas (eDIG) around CHANG-ES galaxies}",
      journal = {\mnras},
         year = 2023,
        month = mar,
       volume = {519},
       number = {4},
        pages = {6098-6110},
          doi = {10.1093/mnras/stad006},
archivePrefix = {arXiv},
       eprint = {2212.14824},
 primaryClass = {astro-ph.GA},
       adsurl = {https://ui.adsabs.harvard.edu/abs/2023MNRAS.519.6098L}
}

@ARTICLE{2012A&ARv..20...49V,
       author = {{Vink}, Jacco},
        title = "{Supernova remnants: the X-ray perspective}",
      journal = {\aapr},
         year = 2012,
        month = dec,
       volume = {20},
          eid = {49},
        pages = {49},
          doi = {10.1007/s00159-011-0049-1},
archivePrefix = {arXiv},
       eprint = {1112.0576},
 primaryClass = {astro-ph.HE},
       adsurl = {https://ui.adsabs.harvard.edu/abs/2012A&ARv..20...49V}
}

@ARTICLE{2024ApJ...967...78L,
       author = {{Li}, Jiang-Tao and {Lu}, Li-Yuan and {Qu}, Zhijie and {Benjamin}, Robert A. and {Bregman}, Joel N. and {Dettmar}, Ralf-J{\"u}rgen and {English}, Jayanne and {Fang}, Taotao and {Irwin}, Judith A. and {Jiang}, Yan and {Li}, Hui and {Liu}, Guilin and {Martini}, Paul and {Rand}, Richard J. and {Stein}, Yelena and {Strong}, Andrew W. and {Vargas}, Carlos J. and {Wang}, Q. Daniel and {Wang}, Jing and {Wiegert}, Theresa and {Xu}, Jianghui and {Yang}, Yang},
        title = "{eDIG-CHANGES. II. Project Design and Initial Results on NGC 3556}",
      journal = {\apj},
         year = 2024,
        month = jun,
       volume = {967},
       number = {2},
          eid = {78},
        pages = {78},
          doi = {10.3847/1538-4357/ad3cd8},
archivePrefix = {arXiv},
       eprint = {2404.05628},
 primaryClass = {astro-ph.GA},
       adsurl = {https://ui.adsabs.harvard.edu/abs/2024ApJ...967...78L}
}

@ARTICLE{2008ApJ...687L..69K,
       author = {{Kenney}, Jeffrey D.~P. and {Tal}, Tomer and {Crowl}, Hugh H. and {Feldmeier}, John and {Jacoby}, George H.},
        title = "{A Spectacular H{\ensuremath{\alpha}} Complex in Virgo: Evidence for a Collision between M86 and NGC 4438 and Implications for the Collisional ISM Heating of Ellipticals}",
      journal = {\apjl},
         year = 2008,
        month = nov,
       volume = {687},
       number = {2},
        pages = {L69},
          doi = {10.1086/593300},
archivePrefix = {arXiv},
       eprint = {0810.0711},
 primaryClass = {astro-ph},
       adsurl = {https://ui.adsabs.harvard.edu/abs/2008ApJ...687L..69K}
}

@ARTICLE{2008Natur.454..968F,
       author = {{Fabian}, A.~C. and {Johnstone}, R.~M. and {Sanders}, J.~S. and {Conselice}, C.~J. and {Crawford}, C.~S. and {Gallagher}, III, J.~S. and {Zweibel}, E.},
        title = "{Magnetic support of the optical emission line filaments in NGC 1275}",
      journal = {\nat},
         year = 2008,
        month = aug,
       volume = {454},
       number = {7207},
        pages = {968-970},
          doi = {10.1038/nature07169},
archivePrefix = {arXiv},
       eprint = {0808.2712},
 primaryClass = {astro-ph},
       adsurl = {https://ui.adsabs.harvard.edu/abs/2008Natur.454..968F}
}

@INPROCEEDINGS{2000ASPC..195...34B,
       author = {{Bland-Hawthorn}, J.},
        title = "{Tunable Imaging Filters and Scientific Applications (Review)}",
    booktitle = {Imaging the Universe in Three Dimensions},
         year = 2000,
       editor = {{van Breugel}, W. and {Bland-Hawthorn}, J.},
       series = {Astronomical Society of the Pacific Conference Series},
       volume = {195},
        month = jan,
        pages = {34},
       adsurl = {https://ui.adsabs.harvard.edu/abs/2000ASPC..195...34B}
}

@INPROCEEDINGS{2022SPIE12188E..21B,
       author = {{Battaini}, Federico and {Ragazzoni}, Roberto and {Milone}, Antonino P. and {Cremonese}, Gabriele},
        title = "{Transmission curves of narrow-band filters in large-FoV and fast astronomical instruments}",
    booktitle = {Advances in Optical and Mechanical Technologies for Telescopes and Instrumentation},
         year = 2022,
       series = {Society of Photo-Optical Instrumentation Engineers (SPIE) Conference Series},
       volume = {12188},
        month = aug,
          eid = {1218821},
        pages = {1218821},
          doi = {10.1117/12.2629936},
archivePrefix = {arXiv},
       eprint = {2211.10150},
 primaryClass = {astro-ph.IM},
       adsurl = {https://ui.adsabs.harvard.edu/abs/2022SPIE12188E..21B}
}

@ARTICLE{2014PASP..126...55A,
       author = {{Abraham}, Roberto G. and {van Dokkum}, Pieter G.},
        title = "{Ultra-Low Surface Brightness Imaging with the Dragonfly Telephoto Array}",
      journal = {\pasp},
         year = 2014,
        month = jan,
       volume = {126},
       number = {935},
        pages = {55},
          doi = {10.1086/674875},
archivePrefix = {arXiv},
       eprint = {1401.5473},
 primaryClass = {astro-ph.IM},
       adsurl = {https://ui.adsabs.harvard.edu/abs/2014PASP..126...55A}
}

@INPROCEEDINGS{2020SPIE11445E..27L,
       author = {{Lokhorst}, Deborah M. and {Abraham}, Roberto G. and {van Dokkum}, Pieter and {Chen}, Seery},
        title = "{Wide-field ultra-narrow-bandpass imaging with the Dragonfly Telephoto Array}",
    booktitle = {Ground-based and Airborne Telescopes VIII},
         year = 2020,
       editor = {{Marshall}, Heather K. and {Spyromilio}, Jason and {Usuda}, Tomonori},
       series = {Society of Photo-Optical Instrumentation Engineers (SPIE) Conference Series},
       volume = {11445},
        month = dec,
          eid = {1144527},
        pages = {1144527},
          doi = {10.1117/12.2561162},
archivePrefix = {arXiv},
       eprint = {2010.00686},
 primaryClass = {astro-ph.IM},
       adsurl = {https://ui.adsabs.harvard.edu/abs/2020SPIE11445E..27L}
}

@ARTICLE{2022ApJ...927..136L,
       author = {{Lokhorst}, Deborah and {Abraham}, Roberto and {Pasha}, Imad and {van Dokkum}, Pieter and {Chen}, Seery and {Miller}, Tim and {Danieli}, Shany and {Greco}, Johnny and {Zhang}, Jielai and {Merritt}, Allison and {Conroy}, Charlie},
        title = "{A Giant Shell of Ionized Gas Discovered near M82 with the Dragonfly Spectral Line Mapper Pathfinder}",
      journal = {\apj},
         year = 2022,
        month = mar,
       volume = {927},
       number = {2},
          eid = {136},
        pages = {136},
          doi = {10.3847/1538-4357/ac50b6},
archivePrefix = {arXiv},
       eprint = {2203.04933},
 primaryClass = {astro-ph.GA},
       adsurl = {https://ui.adsabs.harvard.edu/abs/2022ApJ...927..136L}
}

@ARTICLE{2025AcASn..66...20Z,
       author = {{Zhuang}, Y.~Y. and {Ping}, Y.~D. and {Zhao}, H.~B.},
        title = "{A Near-Earth Object Survey Planning Model for Multi-Application Survey Telescope Array}",
      journal = {Acta Astronomica Sinica},
         year = 2025,
        month = mar,
       volume = {66},
       number = {2},
          eid = {20},
        pages = {20},
          doi = {10.15940/j.cnki.0001-5245.2025.02.009},
       adsurl = {https://ui.adsabs.harvard.edu/abs/2025AcASn..66...20Z}
}

@ARTICLE{2025AcASn..66...58X,
       author = {{Xu}, X. and {Lou}, Z. and {Ping}, Y.~D. and {Dong}, Y.~F. and {Zheng}, X.~Z.},
        title = "{Performance of Narrow-band Imaging for MASTA and WFST}",
      journal = {Acta Astronomica Sinica},
         year = 2025,
        month = sep,
       volume = {66},
       number = {5},
          eid = {58},
        pages = {58},
          doi = {10.15940/j.cnki.0001-5245.2025.05.012},
       adsurl = {https://ui.adsabs.harvard.edu/abs/2025AcASn..66...58X}
}

@misc{ccdproc,
  author = {Craig, Matt and Crawford, Steve and Seifert, Michael and others},
  title = {astropy/ccdproc: 2.5.1: The final 2.x release},
  year = {2025},
  publisher = {Zenodo},
  doi = {10.5281/zenodo.16755473},
  url = {https://doi.org/10.5281/zenodo.16755473}
}

@ARTICLE{1996A&AS..117..393B,
       author = {{Bertin}, E. and {Arnouts}, S.},
        title = "{SExtractor: Software for source extraction.}",
      journal = {\aaps},
         year = 1996,
        month = jun,
       volume = {117},
        pages = {393-404},
          doi = {10.1051/aas:1996164},
       adsurl = {https://ui.adsabs.harvard.edu/abs/1996A&AS..117..393B}
}

@ARTICLE{2010AJ....139.1782L,
       author = {{Lang}, Dustin and {Hogg}, David W. and {Mierle}, Keir and {Blanton}, Michael and {Roweis}, Sam},
        title = "{Astrometry.net: Blind Astrometric Calibration of Arbitrary Astronomical Images}",
      journal = {\aj},
         year = 2010,
        month = may,
       volume = {139},
       number = {5},
        pages = {1782-1800},
          doi = {10.1088/0004-6256/139/5/1782},
archivePrefix = {arXiv},
       eprint = {0910.2233},
 primaryClass = {astro-ph.IM},
       adsurl = {https://ui.adsabs.harvard.edu/abs/2010AJ....139.1782L}
}

@ARTICLE{2025A&A...699A..50G,
       author = {{Garrappa}, S. and {Ofek}, E.~O. and {Ben-Ami}, S. and {Polishook}, D. and {Gal-Yam}, A. and {Shvartzvald}, Y. and {Krassilchtchikov}, A. and {Konno}, R. and {Segre}, E. and {Shani}, Y.~M. and {Sofer-Rimalt}, Y. and {Engel}, M. and {Blumenzweig}, A.},
        title = "{Accurate photometric calibration by fitting the system transmission}",
      journal = {\aap},
         year = 2025,
        month = jul,
       volume = {699},
          eid = {A50},
        pages = {A50},
          doi = {10.1051/0004-6361/202453440},
archivePrefix = {arXiv},
       eprint = {2412.13257},
 primaryClass = {astro-ph.IM},
       adsurl = {https://ui.adsabs.harvard.edu/abs/2025A&A...699A..50G}
}

@ARTICLE{2016A&A...595A...1G,
       author = {{Gaia Collaboration} and {Prusti}, T. and {de Bruijne}, J.~H.~J. and {Brown}, A.~G.~A. and {Vallenari}, A. and {Babusiaux}, C. and {Bailer-Jones}, C.~A.~L. and {Bastian}, U. and {Biermann}, M. and {Evans}, D.~W. and {Eyer}, L. and {Jansen}, F. and {Jordi}, C. and {Klioner}, S.~A. and {Lammers}, U. and {Lindegren}, L. and {Luri}, X. and {Mignard}, F. and {Milligan}, D.~J. and {Panem}, C. and {Poinsignon}, V. and {Pourbaix}, D. and {Randich}, S. and {Sarri}, G. and {Sartoretti}, P. and {Siddiqui}, H.~I. and {Soubiran}, C. and {Valette}, V. and {van Leeuwen}, F. and {Walton}, N.~A. and {Aerts}, C. and {Arenou}, F. and {Cropper}, M. and {Drimmel}, R. and {H{\o}g}, E. and {Katz}, D. and {Lattanzi}, M.~G. and {O'Mullane}, W. and {Grebel}, E.~K. and {Holland}, A.~D. and {Huc}, C. and {Passot}, X. and {Bramante}, L. and {Cacciari}, C. and {Casta{\~n}eda}, J. and {Chaoul}, L. and {Cheek}, N. and {De Angeli}, F. and {Fabricius}, C. and {Guerra}, R. and {Hern{\'a}ndez}, J. and {Jean-Antoine-Piccolo}, A. and {Masana}, E. and {Messineo}, R. and {Mowlavi}, N. and {Nienartowicz}, K. and {Ord{\'o}{\~n}ez-Blanco}, D. and {Panuzzo}, P. and {Portell}, J. and {Richards}, P.~J. and {Riello}, M. and {Seabroke}, G.~M. and {Tanga}, P. and {Th{\'e}venin}, F. and {Torra}, J. and {Els}, S.~G. and {Gracia-Abril}, G. and {Comoretto}, G. and {Garcia-Reinaldos}, M. and {Lock}, T. and {Mercier}, E. and {Altmann}, M. and {Andrae}, R. and {Astraatmadja}, T.~L. and {Bellas-Velidis}, I. and {Benson}, K. and {Berthier}, J. and {Blomme}, R. and {Busso}, G. and {Carry}, B. and {Cellino}, A. and {Clementini}, G. and {Cowell}, S. and {Creevey}, O. and {Cuypers}, J. and {Davidson}, M. and {De Ridder}, J. and {de Torres}, A. and {Delchambre}, L. and {Dell'Oro}, A. and {Ducourant}, C. and {Fr{\'e}mat}, Y. and {Garc{\'\i}a-Torres}, M. and {Gosset}, E. and {Halbwachs}, J.-L. and {Hambly}, N.~C. and {Harrison}, D.~L. and {Hauser}, M. and {Hestroffer}, D. and {Hodgkin}, S.~T. and {Huckle}, H.~E. and {Hutton}, A. and {Jasniewicz}, G. and {Jordan}, S. and {Kontizas}, M. and {Korn}, A.~J. and {Lanzafame}, A.~C. and {Manteiga}, M. and {Moitinho}, A. and {Muinonen}, K. and {Osinde}, J. and {Pancino}, E. and {Pauwels}, T. and {Petit}, J.-M. and {Recio-Blanco}, A. and {Robin}, A.~C. and {Sarro}, L.~M. and {Siopis}, C. and {Smith}, M. and {Smith}, K.~W. and {Sozzetti}, A. and {Thuillot}, W. and {van Reeven}, W. and {Viala}, Y. and {Abbas}, U. and {Abreu Aramburu}, A. and {Accart}, S. and {Aguado}, J.~J. and {Allan}, P.~M. and {Allasia}, W. and {Altavilla}, G. and {{\'A}lvarez}, M.~A. and {Alves}, J. and {Anderson}, R.~I. and {Andrei}, A.~H. and {Anglada Varela}, E. and {Antiche}, E. and {Antoja}, T. and {Ant{\'o}n}, S. and {Arcay}, B. and {Atzei}, A. and {Ayache}, L. and {Bach}, N. and {Baker}, S.~G. and {Balaguer-N{\'u}{\~n}ez}, L. and {Barache}, C. and {Barata}, C. and {Barbier}, A. and {Barblan}, F. and {Baroni}, M. and {Barrado y Navascu{\'e}s}, D. and {Barros}, M. and {Barstow}, M.~A. and {Becciani}, U. and {Bellazzini}, M. and {Bellei}, G. and {Bello Garc{\'\i}a}, A. and {Belokurov}, V. and {Bendjoya}, P. and {Berihuete}, A. and {Bianchi}, L. and {Bienaym{\'e}}, O. and {Billebaud}, F. and {Blagorodnova}, N. and {Blanco-Cuaresma}, S. and {Boch}, T. and {Bombrun}, A. and {Borrachero}, R. and {Bouquillon}, S. and {Bourda}, G. and {Bouy}, H. and {Bragaglia}, A. and {Breddels}, M.~A. and {Brouillet}, N. and {Br{\"u}semeister}, T. and {Bucciarelli}, B. and {Budnik}, F. and {Burgess}, P. and {Burgon}, R. and {Burlacu}, A. and {Busonero}, D. and {Buzzi}, R. and {Caffau}, E. and {Cambras}, J. and {Campbell}, H. and {Cancelliere}, R. and {Cantat-Gaudin}, T. and {Carlucci}, T. and {Carrasco}, J.~M. and {Castellani}, M. and {Charlot}, P. and {Charnas}, J. and {Charvet}, P. and {Chassat}, F. and {Chiavassa}, A. and {Clotet}, M. and {Cocozza}, G. and {Collins}, R.~S. and {Collins}, P. and {Costigan}, G.},
        title = "{The Gaia mission}",
      journal = {\aap},
         year = 2016,
        month = nov,
       volume = {595},
          eid = {A1},
        pages = {A1},
          doi = {10.1051/0004-6361/201629272},
archivePrefix = {arXiv},
       eprint = {1609.04153},
 primaryClass = {astro-ph.IM},
       adsurl = {https://ui.adsabs.harvard.edu/abs/2016A&A...595A...1G}
}

@ARTICLE{2023A&A...674A...1G,
       author = {{Gaia Collaboration} and {Vallenari}, A. and {Brown}, A.~G.~A. and {Prusti}, T. and {de Bruijne}, J.~H.~J. and {Arenou}, F. and {Babusiaux}, C. and {Biermann}, M. and {Creevey}, O.~L. and {Ducourant}, C. and {Evans}, D.~W. and {Eyer}, L. and {Guerra}, R. and {Hutton}, A. and {Jordi}, C. and {Klioner}, S.~A. and {Lammers}, U.~L. and {Lindegren}, L. and {Luri}, X. and {Mignard}, F. and {Panem}, C. and {Pourbaix}, D. and {Randich}, S. and {Sartoretti}, P. and {Soubiran}, C. and {Tanga}, P. and {Walton}, N.~A. and {Bailer-Jones}, C.~A.~L. and {Bastian}, U. and {Drimmel}, R. and {Jansen}, F. and {Katz}, D. and {Lattanzi}, M.~G. and {van Leeuwen}, F. and {Bakker}, J. and {Cacciari}, C. and {Casta{\~n}eda}, J. and {De Angeli}, F. and {Fabricius}, C. and {Fouesneau}, M. and {Fr{\'e}mat}, Y. and {Galluccio}, L. and {Guerrier}, A. and {Heiter}, U. and {Masana}, E. and {Messineo}, R. and {Mowlavi}, N. and {Nicolas}, C. and {Nienartowicz}, K. and {Pailler}, F. and {Panuzzo}, P. and {Riclet}, F. and {Roux}, W. and {Seabroke}, G.~M. and {Sordo}, R. and {Th{\'e}venin}, F. and {Gracia-Abril}, G. and {Portell}, J. and {Teyssier}, D. and {Altmann}, M. and {Andrae}, R. and {Audard}, M. and {Bellas-Velidis}, I. and {Benson}, K. and {Berthier}, J. and {Blomme}, R. and {Burgess}, P.~W. and {Busonero}, D. and {Busso}, G. and {C{\'a}novas}, H. and {Carry}, B. and {Cellino}, A. and {Cheek}, N. and {Clementini}, G. and {Damerdji}, Y. and {Davidson}, M. and {de Teodoro}, P. and {Nu{\~n}ez Campos}, M. and {Delchambre}, L. and {Dell'Oro}, A. and {Esquej}, P. and {Fern{\'a}ndez-Hern{\'a}ndez}, J. and {Fraile}, E. and {Garabato}, D. and {Garc{\'\i}a-Lario}, P. and {Gosset}, E. and {Haigron}, R. and {Halbwachs}, J.-L. and {Hambly}, N.~C. and {Harrison}, D.~L. and {Hern{\'a}ndez}, J. and {Hestroffer}, D. and {Hodgkin}, S.~T. and {Holl}, B. and {Jan{\ss}en}, K. and {Jevardat de Fombelle}, G. and {Jordan}, S. and {Krone-Martins}, A. and {Lanzafame}, A.~C. and {L{\"o}ffler}, W. and {Marchal}, O. and {Marrese}, P.~M. and {Moitinho}, A. and {Muinonen}, K. and {Osborne}, P. and {Pancino}, E. and {Pauwels}, T. and {Recio-Blanco}, A. and {Reyl{\'e}}, C. and {Riello}, M. and {Rimoldini}, L. and {Roegiers}, T. and {Rybizki}, J. and {Sarro}, L.~M. and {Siopis}, C. and {Smith}, M. and {Sozzetti}, A. and {Utrilla}, E. and {van Leeuwen}, M. and {Abbas}, U. and {{\'A}brah{\'a}m}, P. and {Abreu Aramburu}, A. and {Aerts}, C. and {Aguado}, J.~J. and {Ajaj}, M. and {Aldea-Montero}, F. and {Altavilla}, G. and {{\'A}lvarez}, M.~A. and {Alves}, J. and {Anders}, F. and {Anderson}, R.~I. and {Anglada Varela}, E. and {Antoja}, T. and {Baines}, D. and {Baker}, S.~G. and {Balaguer-N{\'u}{\~n}ez}, L. and {Balbinot}, E. and {Balog}, Z. and {Barache}, C. and {Barbato}, D. and {Barros}, M. and {Barstow}, M.~A. and {Bartolom{\'e}}, S. and {Bassilana}, J.-L. and {Bauchet}, N. and {Becciani}, U. and {Bellazzini}, M. and {Berihuete}, A. and {Bernet}, M. and {Bertone}, S. and {Bianchi}, L. and {Binnenfeld}, A. and {Blanco-Cuaresma}, S. and {Blazere}, A. and {Boch}, T. and {Bombrun}, A. and {Bossini}, D. and {Bouquillon}, S. and {Bragaglia}, A. and {Bramante}, L. and {Breedt}, E. and {Bressan}, A. and {Brouillet}, N. and {Brugaletta}, E. and {Bucciarelli}, B. and {Burlacu}, A. and {Butkevich}, A.~G. and {Buzzi}, R. and {Caffau}, E. and {Cancelliere}, R. and {Cantat-Gaudin}, T. and {Carballo}, R. and {Carlucci}, T. and {Carnerero}, M.~I. and {Carrasco}, J.~M. and {Casamiquela}, L. and {Castellani}, M. and {Castro-Ginard}, A. and {Chaoul}, L. and {Charlot}, P. and {Chemin}, L. and {Chiaramida}, V. and {Chiavassa}, A. and {Chornay}, N. and {Comoretto}, G. and {Contursi}, G. and {Cooper}, W.~J. and {Cornez}, T. and {Cowell}, S. and {Crifo}, F. and {Cropper}, M. and {Crosta}, M. and {Crowley}, C. and {Dafonte}, C. and {Dapergolas}, A. and {David}, M. and {David}, P. and {de Laverny}, P. and {De Luise}, F. and {De March}, R.},
        title = "{Gaia Data Release 3. Summary of the content and survey properties}",
      journal = {\aap},
         year = 2023,
        month = jun,
       volume = {674},
          eid = {A1},
        pages = {A1},
          doi = {10.1051/0004-6361/202243940},
archivePrefix = {arXiv},
       eprint = {2208.00211},
 primaryClass = {astro-ph.GA},
       adsurl = {https://ui.adsabs.harvard.edu/abs/2023A&A...674A...1G}
}

@ARTICLE{1993A&A...268..419C,
       author = {{Courtes}, G. and {Petit}, H. and {Hua}, C.~T. and {Martin}, P. and {Blecha}, A. and {Huguenin}, D. and {Golay}, M.},
        title = "{Structure of the spiral arms of NGC 4258 in H-alpha and at 2000A}",
      journal = {\aap},
         year = 1993,
        month = feb,
       volume = {268},
        pages = {419-442},
       adsurl = {https://ui.adsabs.harvard.edu/abs/1993A&A...268..419C}
}

@ARTICLE{2025MNRAS.542...25F,
       author = {{Fern{\'a}ndez-Arenas}, D. and {Rousseau-Nepton}, L. and {Robert}, C. and {Drissen}, L. and {Martin}, R.~P. and {Amram}, P. and {Epinat}, B. and {Duarte Puertas}, S. and {Garner}, III, R. and {Savard}, G. and {Vicens-Mouret}, S. and {Mass{\'e}}, {\'E}.},
        title = "{Ionized gas in NGC 4258: exploring the AGN-star formation connection}",
      journal = {\mnras},
         year = 2025,
        month = sep,
       volume = {542},
       number = {1},
        pages = {25-51},
          doi = {10.1093/mnras/staf1177},
archivePrefix = {arXiv},
       eprint = {2507.16153},
 primaryClass = {astro-ph.GA},
       adsurl = {https://ui.adsabs.harvard.edu/abs/2025MNRAS.542...25F}
}

@ARTICLE{2010MNRAS.406..181J,
       author = {{Jim{\'e}nez-Vicente}, J. and {Mediavilla}, E. and {Castillo-Morales}, A. and {Battaner}, E.},
        title = "{Blasts and shocks in the disc of NGC 4258}",
      journal = {\mnras},
         year = 2010,
        month = jul,
       volume = {406},
       number = {1},
        pages = {181-186},
          doi = {10.1111/j.1365-2966.2010.16692.x},
archivePrefix = {arXiv},
       eprint = {1003.3635},
 primaryClass = {astro-ph.GA},
       adsurl = {https://ui.adsabs.harvard.edu/abs/2010MNRAS.406..181J}
}

\label{lastpage}

\end{document}